\documentclass[aps,prb,twocolumn,superscriptaddress,a4paper]{revtex4-1}
\usepackage[dvipdfmx]{graphicx}

\usepackage{amsmath,amsthm,amssymb}
\usepackage{amsmath,bm}
\usepackage{color}
\usepackage{ifthen}
\usepackage{mathrsfs}

\usepackage{multirow,bigdelim}

\newcount \commentout
\newcommand{\1}{\mbox{1}\hspace{-0.25em}\mbox{l}}

\def\qqqqquad{\quad\quad\quad\quad\quad}
\def\tenquad{\quad\quad\quad\quad\quad\quad\quad\quad\quad\quad}

\newcommand{\blue}[1]{{{#1}}}

\newcommand{\sgn}{\mathrm{sgn}}

\def\rrangle{\rangle\!\rangle}
\def\llangle{\langle\!\langle}

\def\spopf#1{\mathscr{#1}}
\def\matf#1{\mathcal{#1}}
\newcommand{\I}{\mathrm{I}}
\newcommand{\II}{\mathrm{I\!I}}

\newlength{\figwidth}
\newlength{\figlarge}
\begin{document}
\title{
Fate of fractional quantum Hall states in open quantum systems: characterization of correlated topological states for the full Liouvillian
}

\author{Tsuneya Yoshida}
\affiliation{
Graduate School of Pure and Applied Sciences, University of Tsukuba, Tsukuba, Ibaraki 305-8571, Japan
}
\affiliation{
Department of Physics, University of Tsukuba, Ibaraki 305-8571, Japan
}
\author{Koji Kudo}
\affiliation{
Graduate School of Pure and Applied Sciences, University of Tsukuba, Tsukuba, Ibaraki 305-8571, Japan
}
\author{Hosho Katsura}
\affiliation{
Department of Physics, Graduate School of Science, The University of Tokyo, Hongo, Tokyo 113-0033, Japan
}
\affiliation{
Institute for Physics of Intelligence, The University of Tokyo, Hongo, Tokyo 113-0033, Japan
}
\affiliation{
Trans-scale Quantum Science Institute, The University of Tokyo, 7-3-1, Hongo, Bunkyo-ku, Tokyo, 113-0033, Japan
}
\author{Yasuhiro Hatsugai}
\affiliation{
Graduate School of Pure and Applied Sciences, University of Tsukuba, Tsukuba, Ibaraki 305-8571, Japan
}
\affiliation{
Department of Physics, University of Tsukuba, Ibaraki 305-8571, Japan
}
\date{\today}
\begin{abstract}
Despite previous extensive analysis of open quantum systems described by the Lindblad equation, it is unclear whether correlated topological states, such as fractional quantum Hall states, are maintained even in the presence of the jump term.
In this paper, we introduce the pseudo-spin Chern number of the Liouvillian which is computed by twisting the boundary conditions only for one of the subspaces of the doubled Hilbert space.
The existence of such a topological invariant elucidates that the topological properties remain unchanged even in the presence of the jump term which does not close the gap of the effective non-Hermitian Hamiltonian (obtained by neglecting the jump term). 
In other words, the topological properties are encoded into an effective non-Hermitian Hamiltonian rather than the full Liouvillian.
This is particularly useful when the jump term can be written as a strictly block-upper (-lower) triangular matrix in the doubled Hilbert space, in which case the presence or absence of the jump term does not affect the spectrum of the Liouvillian. 
With the pseudo-spin Chern number, we address the characterization of fractional quantum Hall states with two-body loss but without gain, elucidating that the topology of the non-Hermitian fractional quantum Hall states is preserved even in the presence of the jump term.
This numerical result also supports the use of the non-Hermitian Hamiltonian which significantly reduces the numerical cost.
Similar topological invariants can be extended to treat correlated topological states for other spatial dimensions and symmetry (e.g., one-dimensional open quantum systems with inversion symmetry), indicating the high versatility of our approach.
\end{abstract}
\maketitle


\section{
Introduction
}
\label{sec: intro}

Recent extensive studies of non-Hermitian systems have discovered a variety of novel topological phenomena for non-interacting cases~\cite{Hu_nHTI_PRB11,Esaki_nHTI_PRB11,Sato_nHTI_PTEP12,Bergholtz_nHReview_arXiv19}. For instance, non-Hermiticity enriches topological properties~\cite{Kawabata_gapped_PRX19}; it increases the number of symmetry classes and results in two types of the gap, the point-gap~\cite{Gong_class_PRX18} and the line-gap~\cite{Shen_nH_band_PRL18}.
Furthermore, non-Hermiticity may break down diagonalizability of the Hamiltonian which results in non-Hermitian band touching~\blue{\cite{BZhen_nH-PHC_Nat15,Shen_nH_band_PRL18,Budich_SPERs_PRB19,Okugawa_SPERs_PRB19,Yoshida_SPERs_PRB19,Zhou_SPERs_Optica19,Kawabata_gapless_class_PRL19,Carlstrom_nHknot_PRA18,Carlstrom_nHknot_PRB18}}, such as exceptional points~\cite{BZhen_nH-PHC_Nat15,Shen_nH_band_PRL18}, symmetry-protected exceptional rings~\cite{Budich_SPERs_PRB19,Okugawa_SPERs_PRB19,Yoshida_SPERs_PRB19,Zhou_SPERs_Optica19,Kawabata_gapless_class_PRL19} etc.
In addition, non-Hermitian systems can also show the intriguing bulk-boundary correspondence~\cite{Alvarez_nHSkin_PRB18,KFlore_nHSkin_PRL18,SYao_nHSkin-1D_PRL18,SYao_nHSkin-2D_PRL18,EElizabet_PRBnHSkinHOTI_PRB19,Rui_Rskin_PRB19,Yokomizo_BBC_PRL19,Xiao_nHSkin_Exp_arXiv19,Kawabata_NBlochBBC_arXiv20}; certain topological properties result in the non-Hermitian skin effect which results in extreme sensitivity to the boundary conditions~\cite{Lee_Skin19,Zhang_BECskin19,Okuma_BECskin19,Yoshida_MSkinarXiv19}. 
So far, the above non-Hermitian phenomena for the non-interacting case have been reported in various platforms~\cite{Guo_nHExp_PRL09,Ruter_nHExp_NatPhys10,Szameit_PRA11,Regensburger_nHExp_Nat12,BZhen_nH-PHC_Nat15,Lee_nHSSH_photo_PRL16,Hassan_EP_PRL17,Feng_nH-PHC_NatPhoto17,Takata_nH-PHC_PRL18,Zhou_BFarc_PHC_Science18,Takata_nH-PHC_OSA19,Ozawa_nHPHC_PMP19,VKozii_nH_arXiv17,Zyuzin_nHEP_PRB18,HShen2018quantum_osci,Yoshida_EP_DMFT_PRB18,Papaji_nHEP_PRB19,Kimura_SPERs_PRB19,Matsushita_ER_PRB19,Yoshida_nHRev_arXiv20,Yoshida_SPERs_mech19,Ghatak_mechSkin_arXiv19,Scheibner_mechEP_arXiv20}.

Among them, open quantum systems~\blue{\cite{Diehl_DissCher_NatPhys11,Bardyn_DissCher_NJP2013,Rivas_DissCher_PRB13,Budich_DissCher_PRA15,Budich_DissCher_PRB15,Xu_ERing3D_PRL17,Goldstein_SciPost19,Shavit_PRB20}} also provide a unique platform of the following intriguing issue: the interplay between correlations and non-Hermitian topology~\cite{Yoshida_nHFQH19,Mu_MBdySkin_arXiv19,Zhang_nHTMI_arXiv20,Liu_nHTMI_arXiv20,Xu_nHBM_arXiv20,Pan_PTHubb_oQS_arXiv20,Lee_corr_1D_Z2_PRB101}.
Such systems interact with the environment and may lose energy or particles. 
Correspondingly, the time-evolution of the density matrix is governed by the Lindblad equation where the coupling between the system and the environment is described by the Lindblad operators $L_\alpha$ ($\alpha=1,2,\cdots$). 
In the previous works~\cite{Yoshida_nHFQH19,Mu_MBdySkin_arXiv19,Zhang_nHTMI_arXiv20,Liu_nHTMI_arXiv20,Xu_nHBM_arXiv20,Pan_PTHubb_oQS_arXiv20,Lee_corr_1D_Z2_PRB101}, by focusing on the special time-evolution, the correlated topological states have been analyzed for the effective non-Hermitian Hamiltonian $H_{\mathrm{eff}}:=H_0-\frac{i}{2}\sum_{\alpha} L^\dagger_\alpha L_\alpha$, where $H_0$ is the Hermitian Hamiltonian of the system; for the short-time dynamics before the occurrence of a jump of the states by Lindblad operators, one can see that the dynamics of the density matrix is described by the effective non-Hermitian Hamiltonian $H_{\mathrm{eff}}$.
Recently, it has been pointed out that for non-interacting fermions, the topological properties can survive even beyond the above special dynamics~\cite{Lieu_Liouclass_PRL20}.
This is because the gap of the Liouvillian is maintained even when the quantum jump is taken into account.

In spite of the above significant progress in topological perspective on open quantum systems, it is still unclear whether the topological properties for correlated states survive even in the presence of quantum jumps. 
In order to clarify the stability of correlated topological phases described by $H_{\mathrm{eff}}$ against the jump term, topological invariants having the following properties should be introduced: (i) they are quantized as long as the gap of the Liouvillian opens; (ii) in the absence of the jump term, they are reduced to the invariants characterizing the topology of the effective non-Hermitian Hamiltonian $H_{\mathrm{eff}}$.

In this paper, to characterize the correlated states, we introduce a topological invariant having the above two properties by doubling the Hilbert space. 
Specifically, we define the pseudo-spin Chern number characterizing the correlated topological states for two-dimensional systems without symmetry~\cite{classA_ftnt}.
This topological invariant can be computed by twisting the boundary conditions for one of the subspaces of the doubled Hilbert space, which is reminiscent of the spin Chern number~\cite{Sheng_spinCh_PRL05,Sheng_spinCh_PRL06,Fukui_spinCh_PRB07}.
By computing the pseudo-spin Chern number, we demonstrate that even in the presence of the jump term, topological properties of non-Hermitian fractional quantum Hall (FQH) states survive for an open quantum system with two-body loss but without gain.
Our results justify the use of the effective non-Hermitian Hamiltonian to topologically characterize the full Liouvillian whose gap does not close even in the presence of the jump term. 
This is particularly useful for systems where the jump term can be written as a block-upper-triangular matrix in the doubled Hilbert space; in such cases, both the spectral and topological properties are encoded into the effective non-Hermitian Hamiltonian which significantly reduces the numerical cost.
We also note that our approach can be extended to characterize correlated topological states for other cases of spatial dimensions and symmetry, indicating the high versatility of our approach.

The rest of this paper is organized as follows.
In Sec.~\ref{sec: rev Lindblad Heff}, we briefly review how the effective non-Hermitian Hamiltonian $H_\mathrm{eff}$ is obtained and provide a detailed description of topological properties which we will discuss in this paper.
In Sec.~\ref{sec: ps Chern}, we introduce the pseudo-spin Chern number of the Liouvillian. 
As an application, we demonstrate that 
for the system with two-body loss but without gain, the topological properties of non-Hermitian FQH states are not affected by the jump term 
in Sec.~\ref{sec: App nHFQH} which is followed by a short summary.
The appendices are devoted to the topological characterization of one-dimensional open quantum systems with inversion symmetry, topological degeneracy for open quantum systems conserving the number of particles, and technical details.

\section{
Effective non-Hermitian Hamiltonian for open quantum systems
}
\label{sec: rev Lindblad Heff}
\subsection{
\blue{
Lindblad equation and the effective non-Hermitian Hamiltonian
}
}

In this section, we briefly review the time-evolution of open quantum systems and concretely explain topological properties on which we will focus in this paper.

Firstly, we note that for open quantum systems, the dynamics is governed by the Lindblad equation,
\begin{subequations}
\begin{eqnarray}
\label{eq: Lindblad}
i\frac{\partial }{\partial t}
\rho&=&
\spopf{L}[\rho]
:=
\spopf{L}_0[\rho]+\spopf{L}_{\mathrm{J}}[\rho],
\end{eqnarray}
where
%
\begin{eqnarray}
\label{eq: L_0}
\spopf{L}_0[\rho]&:=& [H_{0}, \rho]-\frac{i}{2} \sum_{\alpha} \left\{ \rho,L^\dagger_\alpha L_\alpha\right\}, \\
\label{eq: L_jump}
\spopf{L}_{\mathrm{J}}[\rho] &:=& i \sum_{\alpha} L_\alpha \rho L^\dagger_\alpha. 
\end{eqnarray}
\end{subequations}
Here, the Lindblad operators are denoted by a set of $L_\alpha$ ($\alpha=1,2,\cdots$) which describes the dissipation arising from coupling to the environment.
The density matrix of the system is denoted by $\rho(t)$.
The superoperator $\spopf{L}[\, \cdot\, ]$ ($\spopf{L}_{\mathrm{J}}[\, \cdot\, ]$) is referred to as the Liouvillian (the jump term). For the details of superoperators, see Appendix~\ref{sec: Choi iso app}.
The operator $H_0$ denotes the Hamiltonian for the system ($H_0=H^\dagger_0$).
\blue{
For arbitrary operators $A$ and $B$, the commutator (anti-commutator) is written as $[A,B]$ ($\{A,B\}$).
}

In some previous works~\cite{Xu_ERing3D_PRL17,Gong_class_PRX18,Yoshida_nHFQH19,Mu_MBdySkin_arXiv19,Zhang_nHTMI_arXiv20,Liu_nHTMI_arXiv20,Xu_nHBM_arXiv20,Pan_PTHubb_oQS_arXiv20} on open quantum systems, topological phenomena have been studied for the effective non-Hermitian Hamiltonian,
\begin{eqnarray}
\label{eq: Heff gen}
H_{\mathrm{eff}}
&=& 
H_0-\frac{i}{2}\sum_{\alpha}L^\dagger_{\alpha}L_{\alpha},
\end{eqnarray}
by focusing on the dynamics before occurrence of a jump of the state by $\spopf{L}_{\mathrm{J}}$, which is described by $i\partial_t \rho(t)=H_{\mathrm{eff}}\rho(t) -\rho(t) H^\dagger_{\mathrm{eff}}$.
For instance, the Chern number $C_{H_{\mathrm{eff}}}$ is computed with the right and left eigenvectors of the non-Hermitian Hamiltonian $H_{\mathrm{eff}}$ for a two-dimensional system without symmetry~\cite{Yoshida_nHFQH19}. 

Here, in order to elucidate effects of the jump term, let us consider the operator $\spopf{L}(\lambda)$ interpolating between $\spopf{L}_0$ and $\spopf{L}_0+\spopf{L}_{\mathrm{J}}$; $\spopf{L}(\lambda):=\spopf{L}_0+\lambda \spopf{L}_{\mathrm{J}}$ ($ 0 \leq \lambda \leq  1$).
With a slight abuse of terminology, we also call $\spopf{L}(\lambda)$ ``Liouvillian"~\blue{\cite{interp_Liouvi_ftnt}}.
When the gap-closing of the ``Liouvillian" $\spopf{L}(\lambda)$ does not occur for an arbitrary value of $\lambda$, the topological properties are expected to be maintained. 
[The gap is defined in Eq.~(\ref{eq: def of gap})].
However, it remains unclear whether there exists a topological invariant that characterizes the topological properties even in the presence of the jump term.

Previous works~\cite{Diehl_DissCher_NatPhys11,Bardyn_DissCher_NJP2013,Rivas_DissCher_PRB13,Budich_DissCher_PRA15,Budich_DissCher_PRB15} have addressed how the presence of the jump term affects the topological characterization of open quantum systems in non-interacting cases.
We note, however, that topological invariants introduced in these previous works can change without the gap-closing in the spectrum of the Liouvillian $\spopf{L}=\spopf{L}_0+\spopf{L}_{\mathrm{J}}$.
For instance, the topological characterizations proposed in Refs.~\onlinecite{Diehl_DissCher_NatPhys11,Bardyn_DissCher_NJP2013,Budich_DissCher_PRA15} require the gap in the spectrum of the density matrix, which is not necessary in our framework.

\subsection{
\blue{
Vectorized density matrices in the doubled Hilbert space
}
}
\label{sec: vectorized rho}
\blue{
For later use, we define ``eigenvalues" and ``eigenvectors" of the Liouvillian $\spopf{L}$ which can be thought of as a non-Hermitian matrix in a doubled Hilbert space, $\mathrm{Ket}\otimes\mathrm{Bra}$. 
}
\blue{
With the following isomorphism, the density matrix is mapped to a vector in the doubled Hilbert space~\cite{Jamiolkowski_Choiiso_RepMathPhys1972,Choi_Choiiso_1975,Verstraete_Choiiso_PRL04,Zwolak_Choiiso_PRL04,Jiang_Choiiso_PRA13,Znidaric_Choiiso_PRE14,Znidaric_Choiiso_PRE15,Prosen_exactHubb_PRL16,Minganti_Choiiso_PRA18,Shibata_Choiiso_PRB19,Yoshioka_Choiiso_PRB19,Ziolkowska_exactHubb_arXiv19,Wolff_Choiiso_arXiv20},
\begin{eqnarray}
\label{eq: choi iso rho}
\rho=\sum_{ij}\rho_{ij} |\phi_i\rangle \langle \phi_j| 
&\leftrightarrow&
|\rho\rrangle=
\sum_{ij}\rho_{ij} |\phi_i \rrangle_{K}  \otimes | \phi_j \rrangle_{B},
\end{eqnarray}
where $|\phi\rangle$'s are states in the original Hilbert space (or Ket space) generated by acting on the vacuum with creation operators in the real space. The coefficient $\rho_{ij}$ is a complex number.
Here, in order to distinguish elements of the doubled Hilbert space from those of the original Hilbert space, we denote a vector in the subspace $\mathrm{Ket}$ ($\mathrm{Bra}$) as $|\phi_i\rrangle_{K(B)}$. 
}

\blue{
The inner product of vectorized matrices, called the Hilbert-Schmidt inner product, is defined as
\begin{eqnarray}
\llangle A | B \rrangle = \mathrm{tr} \left(A^{\dagger}B\right):= \sum_{ij}A^\dagger_{ij}B_{ji}.
\end{eqnarray}
%
}

\blue{
With the above isomorphism, $L_\alpha \rho L^\dagger_\alpha$ is represented as $L_\alpha \otimes L^*_\alpha |\rho\rrangle$.
Therefore, the Liouvillian $\spopf{L}$ can be represented as a non-Hermitian matrix $\matf{L}$ whose left and right eigenvectors ${}_L\llangle \rho_n|$ and $|\rho_n\rrangle_R$ are defined as
\begin{eqnarray}
\label{eq: rho eigenvec}
\matf{L} |\rho_n\rrangle_R = |\rho_n\rrangle_R \Lambda_n, &\quad& {}_L\llangle \rho_n| \matf{L} = \Lambda_n {}_L\llangle \rho_n|,
\end{eqnarray}
with the eigenvalues $\Lambda_n$, $n=1,2,\cdots$, (for more details, see Appendix~\ref{sec: Choi iso app}). 
}
\blue{
The gap between eigenstates $|\rho_n\rrangle_{R}$ and $|\rho_{n'}\rrangle_{R}$ can be defined as~\cite{gap_ftnt}
\begin{eqnarray}
\label{eq: def of gap}
\Delta &=& \mathrm{Im}(\Lambda_n-\Lambda_{n'}).
\end{eqnarray}
}

\blue{
By $\matf{L}(\lambda):=\matf{L}_0+\lambda \matf{L}_{\mathrm{J}}$, we denote the ``Liouvillian" in the doubled Hilbert space 
which interpolates between the two cases, $\matf{L}(0)=\matf{L}_0= H_{\mathrm{eff}}\otimes \1 -\1\otimes H^*_{\mathrm{eff}} $ and $\matf{L}(1)=\matf{L}=\matf{L}_0+\matf{L}_{\mathrm{J}}$ with $\matf{L}_{\mathrm{J}}=i\sum_{\alpha} L_\alpha \otimes L^*_\alpha$.
}

\section{
Pseudo-spin Chern number for the Liouvillian
}
\label{sec: ps Chern}

In order to clarify whether the topological properties for $H_{\mathrm{eff}}$ are maintained even in the presence of the jump term, we introduce the pseudo-spin Chern number for two-dimensional systems without symmetry.

We note that our approach can be extended to characterize correlated topological states for other spatial dimensions and symmetry [e.g., one-dimensional systems with inversion symmetry, (see Appendix~\ref{sec: 1D L Berry app})], although we limit our discussion to the Chern number for the sake of concreteness.

\subsection{
Definition
}
%
\blue{
Suppose that the gap of the ``Liouvillian" $\matf{L}(\lambda)$ is maintained for $ 0 \leq \lambda \leq  1$ (in the case of topological ordered states~\cite{topo_deg_defs_ftnt}, also suppose that the topological degeneracy is maintained, i.e., the above gap separates the topologically degenerate states from the others), the topological properties are considered to be maintained which are characterized by the topological invariant computed from the eigenvectors of $H_{\mathrm{eff}}$ for $\lambda=0$.
}

The above topological properties can be characterized by the pseudo-spin Chern number $C_{\mathrm{ps}}=(C_{KK}-C_{BB})/2$ where $C_{\sigma\sigma}$ ($\sigma=K,B$) is defined as
\begin{subequations}
\label{eq: Ckk w jump}
\begin{eqnarray}
C_{\sigma\sigma}&:=& \int \frac{d\theta_xd\theta_y}{ 2\pi } \, \mathrm{Im} F_{\sigma\sigma}(\theta_x,\theta_y), \\
F_{\sigma\sigma}&:=& \epsilon_{\mu\nu} \sum_n {}_L\llangle \partial^\sigma_\mu \rho_n| \partial^\sigma_\nu \rho_n \rrangle_{R}.
\end{eqnarray}
\end{subequations}
Here, the summation $\displaystyle{\sum_n}$ is taken over degenerate states;
\blue{
we have supposed that the eigenvectors of the ``Liouvillian" $\matf{L}(\lambda)$ shows $N^2_d$-fold degeneracy for arbitrary $\lambda$, which means that the eigenstates of $H_{\mathrm{eff}}$ show the $N_d$-fold degeneracy [such degeneracy is indeed observed for FQH states with two-body loss (Sec.~\ref{sec: FQH L0})].
}
\blue{The symbol $\epsilon_{\mu\nu}$ denotes the anti-symmetric tensor with $\epsilon_{xy}=-\epsilon_{yx}=1$.}
The summation is taken for repeated indices $\mu$ and $\nu$ [$\mu(\nu)=x,y$].
\blue{
Vectors $|\rho_n \rrangle_R$ and ${}_L \llangle \rho_n|$ are right and left eigenvectors of $\matf{L}(\lambda)$ [see Eq.~(\ref{eq: rho eigenvec})] which 
}
satisfy the biorthogonal normalization condition; $|\rho_n\rrangle_R$ and ${}_L\llangle \rho_{n'}|$, satisfy ${}_L \llangle \rho_{n'}|\rho_n\rrangle_R=\delta_{n'n}$ for arbitrary integers, $n$ and $n'$.
In addition, we have imposed the twisted boundary conditions with  $(\theta_x,\theta_y)$ only for the space specified by $\sigma$~\cite{Niu_HallCond_PRB85,Sheng_spinCh_PRL06,Kudo_psChen_JPSJ18}.
The periodic boundary conditions are imposed on the other space.
The operator $\partial^\sigma_\mu$ denotes the corresponding differential operator acting only on the space specified by $\sigma$.
For instance, the action of $\partial^{K}_\mu$ on a state $|\Phi\rrangle_{K}\otimes|\Phi'\rrangle_{B}$ reads $\left(\partial^{K}_\mu|\Phi\rrangle_{K}\right)\otimes|\Phi'\rrangle_{B}$.

As proven in Sec.~\ref{sec: properties of Cps}, the pseudo-spin Chern number $C_{\mathrm{ps}}$ elucidates that as long as the gap of the ``Liouvillian" $\matf{L}(\lambda)$ opens, the topological properties of $H_{\mathrm{eff}}$ are maintained even in the presence of the jump term.
We note that when the pseudo-spin Chern number changes, the gap-closing should occur in the parameter space of $(\theta_x,\theta_y)$.

The effective non-Hermitian Hamiltonian $H_{\mathrm{eff}}$ is particularly useful when $\mathcal{L}_{\mathrm{J}}$ and $\mathcal{L}_{0}$ can be written in block-upper-triangular and block-diagonal forms, respectively. 
This is because in such cases, the effective non-Hermitian Hamiltonian governs not only topological properties but also spectrum of the full Liouvillian~\cite{Torres_Utrian_PRA14,Nakagawa_exactHubb_arXiv20} (see Appendix~\ref{sec: triangular app}), which significantly reduces the numerical cost.

\subsection{
Properties of the pseudo-spin Chern number
}
\label{sec: properties of Cps}
The pseudo-spin Chern number elucidates that even in the presence of the jump term, topological properties of $H_{\mathrm{eff}}$ remain unchanged as long as the gap of the ``Liouvillian" $\matf{L}(\lambda)$ opens.
In order to see this, we note the following three facts.

(i) The pseudo-spin Chern number is quantized even in the presence of the jump term, provided that the gap-closing of $\matf{L}$ does not occur in the space of $(\theta_x,\theta_y)$. 
\blue{
The quantization of $C_{\sigma\sigma}$ can be proven by extending the argument in Refs.~\onlinecite{Niu_HallCond_PRB85,Kohmoto_AnnPhys1985} (for more details, see Appendix~\ref{sec: quantize ps Ch app}).
We note that introducing a perturbation does not change $C_{\mathrm{ps}}$ as long as the gap is open. 
This can be seen by noting that $C_{\sigma\sigma}$ under the gap condition is continuous as a function of the strength of the perturbation, while its value is quantized.
}

(ii) In the absence of the jump term, 
$C_{KK}$ is rewritten as
\begin{eqnarray}
\label{eq: Ckk to state Ch}
C_{KK}&=& N_d C_{H_{\mathrm{eff}}},
\end{eqnarray}
with
\begin{subequations}
\label{eq: state Ch}
\begin{eqnarray}
C_{H_{\mathrm{eff}}} &=&\int \frac{d\theta_xd\theta_y}{2\pi } \mathrm{Im} f(\theta_x,\theta_y), \\
f(\theta_x,\theta_y)&=& \epsilon_{\mu\nu} \sum_{n_1} {}_L\langle \partial_\mu \Phi_{n_1} | \partial_\nu \Phi_{n_1}\rangle_R.
\end{eqnarray}
\end{subequations}
Equation~(\ref{eq: Ckk to state Ch}) is proven in Sec.~\ref{sec: proof Ckk = state Ch app}.
We note that $C_{H_{\mathrm{eff}}}$ defined in Eq.~(\ref{eq: state Ch}) is nothing but the Chern number of $H_{\mathrm{eff}}$~\cite{Yoshida_nHFQH19}.

(iii) In the absence of the jump term, the Chern number obtained by twisting the boundary conditions only for the subspace $\mathrm{Bra}$ ($C_{BB}$) satisfies,
\begin{eqnarray}
\label{eq: Ckk = -Cbb}
C_{BB} &=& -C_{KK},
\end{eqnarray}
%
which is proven in Sec.~\ref{sec: proof Ckk = -Cbb app}.
This relation also indicates that for $\lambda=0$, the total Chern number computed by twisting the boundary conditions both for the subspaces $\mathrm{Bra}$ and $\mathrm{Ket}$ \blue{(i.e., $C_{\mathrm{tot}}=C_{KK}+C_{BB}$)} vanishes even when the eigenstates of $H_{\mathrm{eff}}$ show topologically non-trivial properties.

Based on the fact (i), we can see that the pseudo-spin Chern number is quantized as long as the gap opens.
In addition, (ii) and (iii) indicate that the pseudo-spin Chern number $C_{\mathrm{ps}}=(C_{KK}-C_{BB})/2$ characterizes the topological properties described by the Hamiltonian $H_{\mathrm{eff}}$ for $\lambda=0$.
Therefore, $C_{\mathrm{ps}}$ elucidates that as long as the gap opens, the topology of $H_{\mathrm{eff}}$ is maintained even in the presence of the jump term.
The effective non-Hermitian Hamiltonian is particularly useful for systems with loss but without gain or vice versa because both the spectral and topological properties are encoded into the effective Hamiltonian $H_{\mathrm{eff}}$ which significantly reduces the numerical cost.

In the rest of this section, we prove Eqs.~(\ref{eq: Ckk to state Ch})~and~(\ref{eq: Ckk = -Cbb}).

\subsubsection{
Proof of Eq.~(\ref{eq: Ckk to state Ch})
}
\label{sec: proof Ckk = state Ch app}
First, we make the identification~\cite{inviso_ftnt}
\begin{eqnarray}
\label{eq: L0 rhoLR}
|\rho_n\rrangle_R \leftrightarrow  |\Phi_{n_1} \rangle_R {}_R \langle \Phi_{n_2} |, &\quad & {}_L \llangle \rho_n | \leftrightarrow |\Phi_{n_2} \rangle_L {}_L \langle \Phi_{n_1} |,\nonumber \\
\end{eqnarray}
where $|\rho_n\rrangle_R$ and ${}_L\llangle \rho_n|$ are right and left eigenvectors of $\matf{L}_0$
\begin{eqnarray}
\matf{L}_0 |\rho_n\rrangle_R &=& (E_{n_1}-E^*_{n_2}) |\rho_n\rrangle_R, \\
 {}_L \llangle \rho_n |\matf{L}_0  &=& {}_L \llangle \rho_n | (E_{n_1}-E^*_{n_2}),
\end{eqnarray}
respectively.
Vectors $|\Phi_{n_1} \rangle_{R}$ and ${}_L\langle \Phi_{n_2} |$ denote the right and left eigenstates of $H_{\mathrm{eff}}$ which satisfy ${}_L\langle \Phi_{n_2} |\Phi_{n_1} \rangle_{R}=\delta_{n_2n_1}$.
The subscript $n$ denotes the set of integers, $n_1$ and $n_2$, labeling the eigenstates, $|\Phi_{n_1} \rangle_{R}$ and ${}_L\langle \Phi_{n_2} |$.

We recall that for the computation of the Chern number $C_{KK}$, the twisted boundary conditions are imposed only on the subspace $\mathrm{Ket}$.
In this case, the derivative $\partial^{K}_\mu$ acts only on the states in the subspace $\mathrm{Ket}$. 
Keeping this fact in mind, we obtain the Berry connection $A_{K\mu}$ and the Berry curvature $F_{KK}$ as
\begin{subequations}
\begin{eqnarray}
A_{K\mu} &:=& \sum_n {}_L \llangle \rho_n | \partial^{K}_\mu | \rho_n \rrangle_R \nonumber \\
         &=& \sum_{n_1n_2} \mathrm{tr} [  | \Phi_{n_2} \rangle_L {}_L \langle \Phi_{n_1}  | \partial_\mu \Phi_{n_1} \rangle_R {}_R \langle \Phi_{n_2} | ] \nonumber \\
         &=& \sum_{n_1n_2} {}_R \langle \Phi_{n_2} | \Phi_{n_2} \rangle_L {}_L \langle \Phi_{n_1}  | \partial_\mu \Phi_{n_1} \rangle_R   \nonumber \\
         &=& N_d\sum_{n_1} {}_L \langle \Phi_{n_1}  | \partial_\mu \Phi_{n_1} \rangle_R,
\end{eqnarray}
and 
\begin{eqnarray}
F_{KK} &:=& \epsilon_{\mu\nu} \partial_\mu A_{K\nu} = N_d\epsilon_{\mu\nu} \sum_{n_1} {}_L\langle \partial_\mu \Phi_{n_1}  | \partial_\nu \Phi_{n_1} \rangle_R. \nonumber \\
\end{eqnarray}
\end{subequations}
Thus, we end up with Eq.~(\ref{eq: Ckk to state Ch}).

\subsubsection{Proof of Eq.~(\ref{eq: Ckk = -Cbb})}
\label{sec: proof Ckk = -Cbb app}
For the computation of the Chern number $C_{BB}$, we impose the twisted boundary conditions only on the subspace $\mathrm{Bra}$, meaning that the derivative $\partial^{B}_\mu$ acts only on the states in the subspace $\mathrm{Bra}$. 
Keeping this in mind, we can see that the Berry connection $A_{B\mu}$ is equal to $A^*_{K\mu}$,
\begin{eqnarray}
\label{eq: Ab=Ak*}
A_{B\mu} &:=& \sum_n {}_L \llangle \rho_n | \partial^{B}_\mu | \rho_n \rrangle_R = N_d \sum_{n_2} {}_R\langle \partial_\mu \Phi_{n_2} | \Phi_{n_2} \rangle_L= A^*_{K\mu}, \nonumber \\
\end{eqnarray}
which yields $F_{BB}:=\epsilon_{\mu\nu} \partial_\mu A_{B\nu}=F^*_{KK}$.

Because the Chern number $C_{BB}$ is an integral of $\mathrm{Im}[F_{BB}]$, we obtain Eq.~(\ref{eq: Ckk = -Cbb}). 

Equation~(\ref{eq: Ab=Ak*}) also indicates that the total Chern number computed by twisting the boundary conditions both for the subspaces $\mathrm{Bra}$ and $\mathrm{Ket}$ \blue{(i.e., $C_{\mathrm{tot}}=C_{KK}+C_{BB}$)} vanishes; 
the Berry connection $A_\mu$ obtained by twisting the boundary conditions both for the subspace satisfies $\mathrm{Im}{A_\mu}=0$, meaning that the relation of $\mathrm{Im}F:=\epsilon_{\mu\nu}\partial_\mu \mathrm{Im}{A_\nu}$ vanishes.

\section{
Application to the FQH states for an open quantum system with two-body loss
}
\label{sec: App nHFQH}
By numerically computing the pseudo-spin Chern number, we elucidate that even in the presence of the jump term, the topology of FQH states survives for the following open quantum system with two-body loss.

\begin{figure}[!h]
\begin{minipage}{0.7\hsize}
\begin{center}
\includegraphics[width=1\hsize,clip]{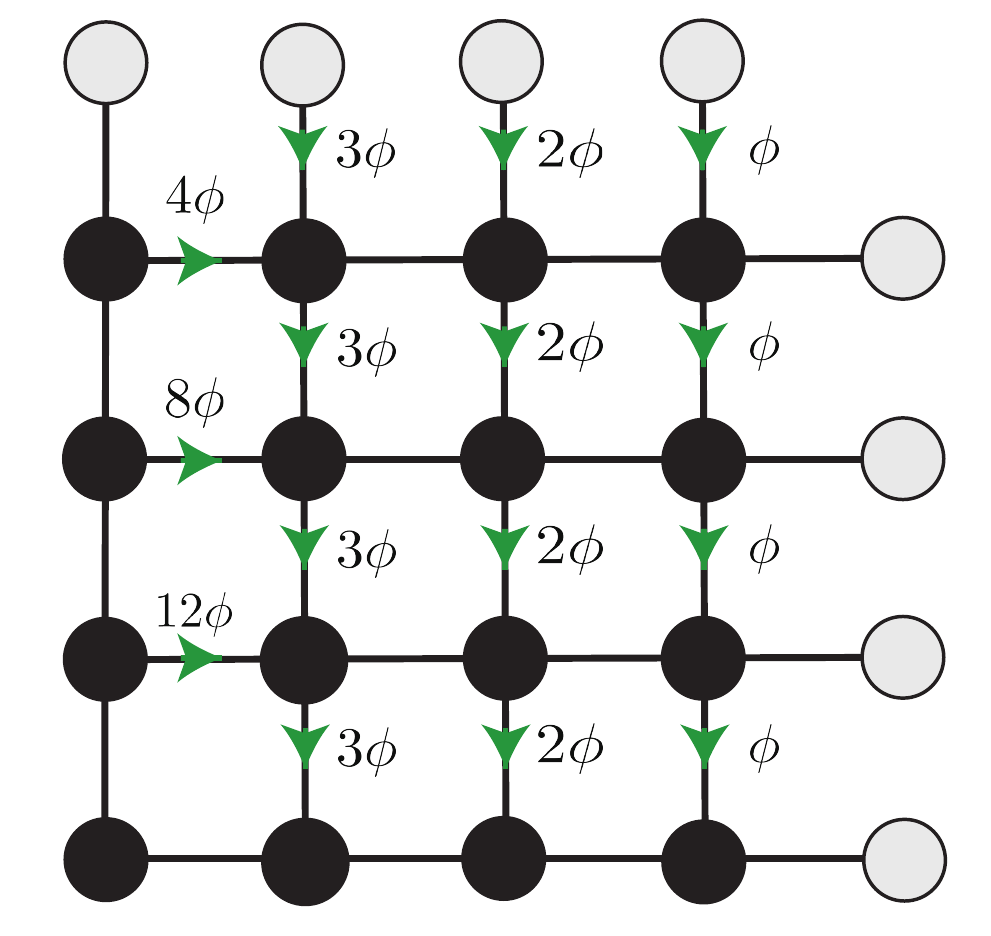}
\end{center}
\end{minipage}
\caption{(Color Online).
Sketch of the model under the periodic boundary conditions. Gray and black circles denote the sites; each site illustrated with a gray circle is identified with the corresponding site illustrated with a black circle on the opposite side. 
To describe the Abelian gauge field, we have taken the string gauge~\cite{Hatsugai_StringG_PRL99}. 
Green arrows illustrate the phase $\phi_{ij}$. 
For hopping parallel to an arrow, $\phi_{ij}$ takes the value shown in the figure. 
When the fermion hops in the opposite direction, $\phi_{ij}$ takes the values so that $\phi_{ij}=-\phi_{ji}$ is satisfied.
The number of the flux quanta penetrating the entire system is written as $N_{\phi}=\phi L_x L_y$ where $L_x$ and $L_y$ denote the number of sites along the $x$- and the $y$-direction, respectively.
When $\phi$ is multiple of $1/L_x$, the string gauge is reduced to the Landau gauge.
}
\label{fig: model}
\end{figure}

Let us consider an open quantum system of spinless fermions on a square lattice. 
We denote by $c^\dagger_i$ and $c_i$ the creation and the annihilation operators of a spinless fermion at site $i$, respectively. 
The number operator at $i$ is defined as $n_i := c^\dagger_i c_i$. 
The system is described by the following Hamiltonian and the Lindblad operators
\begin{subequations}
\label{eq: H0 and Lmu 2bdy}
\begin{eqnarray}
 H_0 &=& \sum_{\langle ij \rangle } h_{ij}c^\dagger_i c_j + V_R \sum_{\langle ij\rangle } n_i n_j, \\
 L_{i\mu} &=& \sqrt{\gamma} c_ic_{i+\bm{e}_\mu},
\end{eqnarray}
\end{subequations}
where $\bm{e}_\mu$ denotes the unit vector in the $\mu$-direction $(\mu=x,y)$. 
The Lindblad operators $L$'s describe two-body loss ($\gamma>0$).
The strength of the nearest neighbor interaction $V_R$ is a real number.
The summation $\displaystyle{\sum_{\langle ij \rangle}}$ is taken over pairs of neighboring sites $i$ and $j$.
The matrix element $h_{ij}=t_0e^{i2\pi\phi_{ij}}$ with real numbers $\phi_{ij}$ and $t_0$ describes hopping between neighboring sites $i$ and $j$ under the gauge field. 
For the definition of the phase factor $\phi_{ij}$, see Fig.~\ref{fig: model} where the string gauge is taken~\cite{Hatsugai_StringG_PRL99}.
The number of the flux quanta penetrating the entire system is written as $N_\phi :=\phi L_x L_y$, where $L_x$ and $L_y$ denote the number of sites along the $x$- and the $y$-direction, respectively.
This model is considered to be relevant to cold atoms.
The Abelian gauge field can be introduced by rotating the system~\cite{Wilkin_fluxRot_PRL98,Schweikhard_fluxRot_PRL04,Sorensen_fluxRot_PRL05,NRCooper_fluxRot_AdvPhys08,Furukawa_fluxRotTheory_PRA12} or by optically synthesized gauge fields~\blue{\cite{Jaksch_SynthGauge_2003,Mueller_SynthGauge_PRA04,Lin_SynthGauge_Nat09,Lin_Syngauge_Nature11,Aidelsburger_Syngauge_PRL13,Miyake_Syngauge_PRL13,Celi_SynthGauge_PRL14,Kessler_Syngauge_PRA14,Jotzu_Syngauge_PRL14,Atala_Syngauge_NatPhys14,Aidelsburger_Syngauge_NatPhys15,Barbarino_SynthGauge_NJP16,Uenal_PRA16,Repellin_PRL19}}.
The Feshbach resonance~\cite{Feshbach_FechbachRes_AnnPhys58,Baumann_FeshbachRes_PRA14} induces inelastic scattering of two-body loss~\cite{Scazza_2bdlossExp_NatPhys14,Pagano_2bdloss_PRL15,Hoefer_2bdlossExp_PRL15,Riegger_2bdlossExp_PRL18,Ashida_nHbHubb_PRA16}.

We address the characterization of non-Hermitian FQH states by the following steps.
Firstly, we rewrite the fermionic open quantum system as a closed fermionic system by identifying the Liouvillian as a non-Hermitian Hamiltonian via the isomorphism [see Eq.~(\ref{eq: choi iso rho})].
Secondly, by numerically diagonalizing the mapped fermionic model~(\ref{eq: C-iso L nHFQH}), we elucidate that the topological properties are maintained; the topological degeneracy and the pseudo-spin Chern number are independent of the jump term.

\subsection{
Mapping the fermionic open quantum system to a closed bilayer system 
}

Firstly, based on the isomorphism [see Eq.~(\ref{eq: choi iso rho})], we show that the systems of spinless fermions with two-body loss can be written as a closed bilayer fermionic system with inter-layer couplings.

With the isomorphism, an annihilation operator $c_{i}$ is mapped to a creation operator $\bar{c}^\dagger_{i}$ for the subspace $\mathrm{Bra}$; $\rho c_{i} \leftrightarrow \bar{c}^\dagger_{i} |\rho \rrangle$ with $\{ \bar{c}_i, \bar{c}^\dagger_{j} \}=\delta_{ij}$ for an arbitrary $\rho$. 
Here, a subtlety arises; commutation relations $[c_{i},\bar{c}_{j}]=[c_{i},\bar{c}^\dagger_{j}]=0$ should hold because the relation $\bar{c}^{\dagger}_{i} |\phi_{j_1} \rrangle_{K}\otimes|\phi_{j_2}\rrangle_{B} = |\phi_{j_1} \rrangle_{K}\otimes\left( \bar{c}^{\dagger}_{i}|\phi_{j_2}\rrangle_{B} \right)$~\cite{cbar_ftnt} holds for arbitrary states $|\phi_{j_1} \rrangle_{K}\otimes|\phi_{j_2}\rrangle_{B}$.

We note, however, that the above commutation relations can be rewritten as the anti-commutation relations by introducing the following operators~\cite{Freericks_PRB95,Bosonization_ftnt}
\begin{eqnarray}
\label{eq: def of dia and dib}
d_{ia} = c_{i},  &\quad& d_{ib} = \bar{c}_{i}P_{fa},
\end{eqnarray}
where $P_{fa}:=(-1)^{\sum_{i}d^\dagger_{ia}d_{ia}}$.
Namely, with the operators $d^\dagger_{i\sigma}$ ($\sigma=a,b$), we have $\{d_{i\sigma},d_{j\sigma'} \}=0$ and $\{d_{i\sigma},d^\dagger_{j\sigma'} \}=\delta_{\sigma\sigma'}\delta_{ij}$.
Here, the operators with $\sigma=a$ ($\sigma=b$) act on the subspace $\mathrm{Ket}$ ($\mathrm{Bra}$).

In terms of the operators $d^\dagger_{i\sigma}$, the Lindblad equation, which is defined with the Hamiltonian $H_0$~(\ref{eq: H0 and Lmu 2bdy}a) and the Lindblad operators~(\ref{eq: H0 and Lmu 2bdy}b), is rewritten as
\begin{subequations}
\label{eq: C-iso L nHFQH}
\begin{eqnarray}
i\partial_t|\rho\rrangle &=& \matf{L}|\rho\rrangle = (\matf{L}_0+ \matf{L}_{\mathrm{J}}) |\rho\rrangle,
\end{eqnarray}
\begin{eqnarray}
\matf{L}_0 &=& \sum_{\langle ij \rangle \sigma} d^\dagger_{i\sigma} h_{ij\sigma} d_{j\sigma} +\sum_{\langle ij\rangle \sigma} V_\sigma n_{i\sigma}n_{j\sigma}, \\
\matf{L}_{\mathrm{J}} &=& -i\gamma\sum_{\langle ij \rangle } d_{ia}d_{ja}d_{jb}d_{ib}, 
\end{eqnarray}
\end{subequations}
with $h_{ija}=h_{ij}$ and $h_{ijb}=-h^*_{ij}$. 
The number operator is defined as $n_{i\sigma}:=d^\dagger_{i\sigma}d_{i\sigma}$.
Here, $V_\sigma=\sgn(\sigma)V_R-i\frac{\gamma}{2}$ with $\sgn(\sigma)$ taking $1$ ($-1$) for $\sigma=a$ ($\sigma=b$).  

The above equation indicates that an open quantum system of spinless fermions can be mapped to a closed bilayer system whose Hamiltonian corresponds to $\matf{L}$ defined in Eq.~(\ref{eq: C-iso L nHFQH}).
Here, we have regarded $d^\dagger_{i\sigma}$ ($\sigma=a,b$) as an operator creating a spinless fermion at site $i$ of layer $\sigma$.

\subsection{
Numerical results
}

\subsubsection{
Overview
}

\begin{figure}[!h]
\begin{minipage}{1.0\hsize}
\begin{center}
\includegraphics[width=1\hsize,clip]{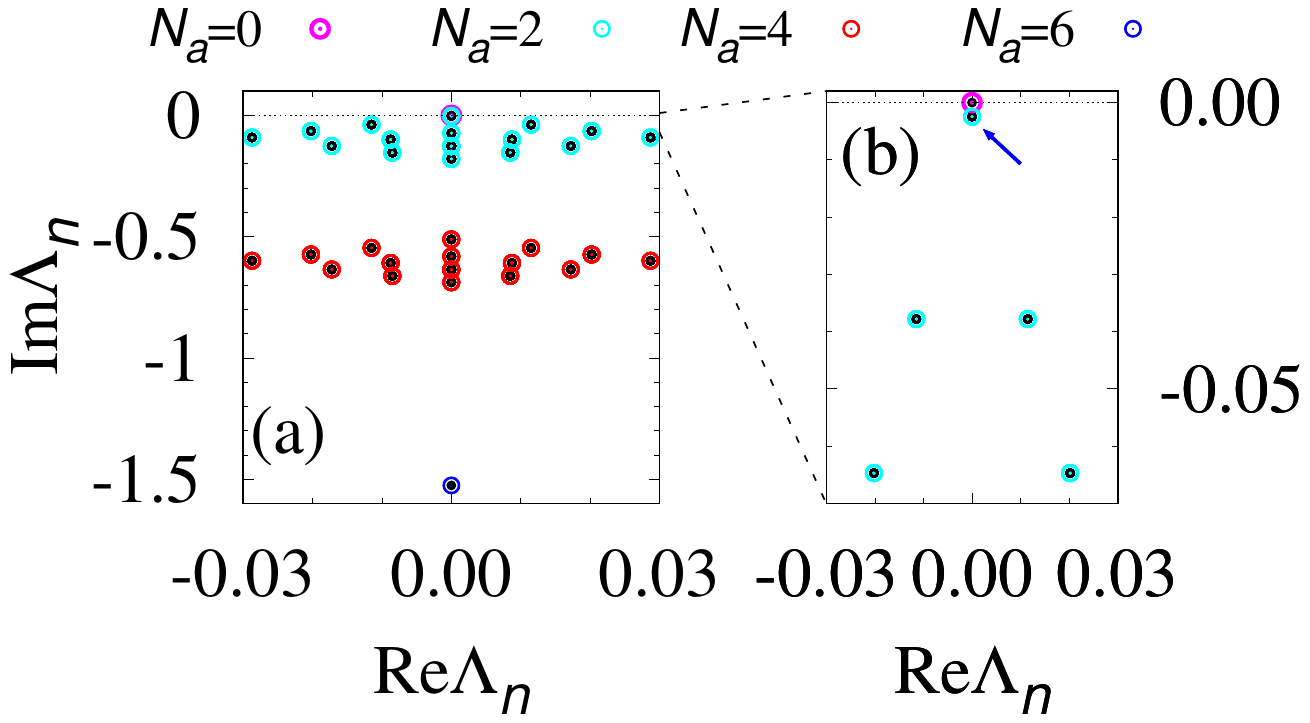}
\end{center}
\end{minipage}
\caption{(Color Online).
Spectrum of the ``Liouvillian" $\matf{L}(\lambda):=\matf{L}_0+\lambda \matf{L}_{\mathrm{J}}$ for $\lambda=0$ (colored dots) and $\lambda=1$ (black dots). 
Explicit forms of $\matf{L}_{0}$ and $\matf{L}_{\mathrm{J}}$ are written in Eq.~(\ref{eq: C-iso L nHFQH}).
The spectra are exactly on top of each other, which is expected from the fact that $\matf{L}_{\mathrm{J}}$ and $\matf{L}_0$ can be written in block-upper-triangular and block-diagonal forms, respectively~\cite{Torres_Utrian_PRA14,Nakagawa_exactHubb_arXiv20} (see Appendix~\ref{sec: triangular app}).
Panel (b) is a magnified version of the range $0\leq \mathrm{Im} \Lambda_n \leq 0.07$ in panel (a).
Parameters are set to $V_R=\cos(0.4\pi)$, $\gamma=2\sin(0.4\pi)$, $t_0=1$, and $L_x=L_y=6$. Total number of flux is $N_\phi=\phi L_xL_y=6$.
The data for $\lambda=0$ (colored dots) are obtained by diagonalizing $\matf{L}_0$ for the subspace labeled by $(N_a,N_b)= (0,0)$, $(2,2)$, $(4,4)$, or $(6,6)$. 
For $(N_a,N_b)=(2,2)$, the filling of each layer is $1/3$.
While the jump term mixes the subspaces labeled by $(N_a,N_b)$ and $(N_a+2,N_b+2)$, the ``Liouvillian" can still be block-diagonalized into subsectors labeled by $\left(N_a-N_b,(-1)^{N_a}\right)$.
The black dots are obtained for the subspace labeled by $\left( N_a-N_b,(-1)^{N_a}\right)=(0,1)$.
The Laughlin states with the filling factor $\nu=1/3$ is denoted by the dots marked with the arrow in panel (b).
We note that the Laughlin states denoted with the arrow has a finite lifetime while the vacuum is a non-equilibrium steady state (i.e, its lifetime is infinite).
}
\label{fig: FQH En}
\end{figure}

\begin{figure}[!h]
\begin{minipage}{0.7\hsize}
\begin{center}
\includegraphics[width=1\hsize,clip]{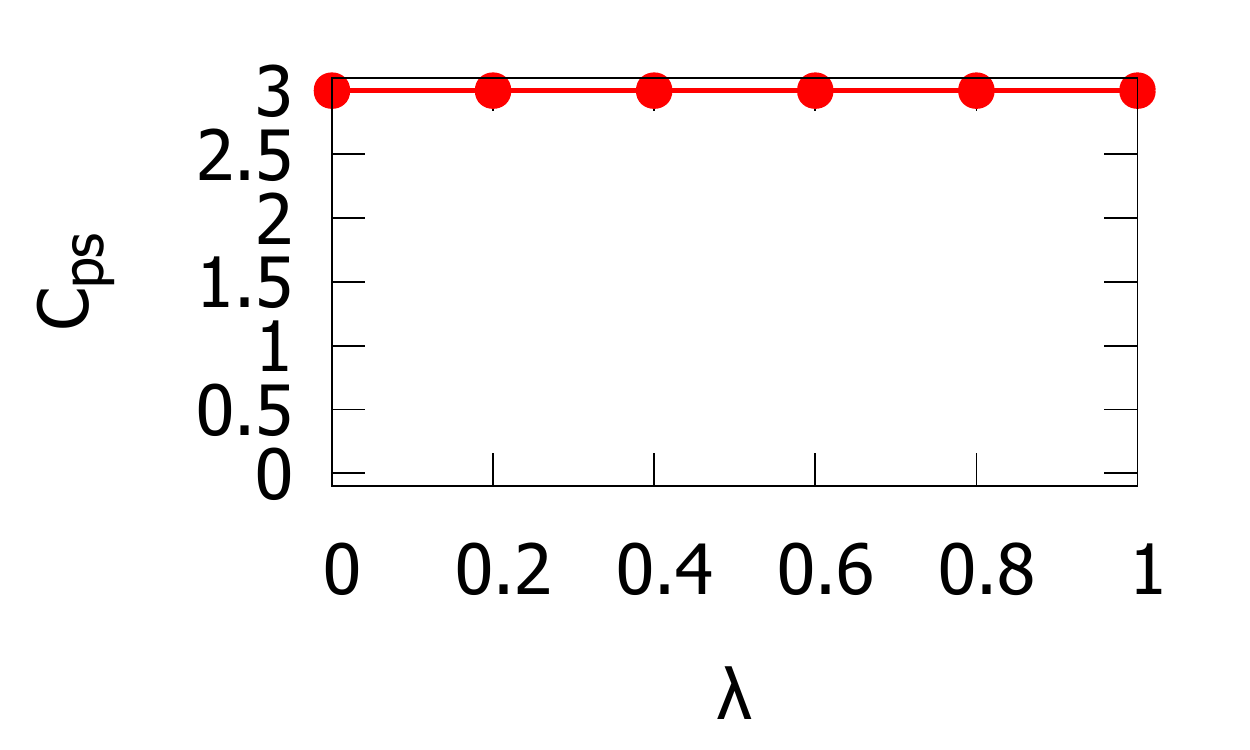}
\end{center}
\end{minipage}
\caption{(Color Online).
The pseudo-spin Chern number as a function of $\lambda$ for the Laughlin states \blue{with 9-fold degeneracy} indicated by dots marked with the blue arrow in Fig.~\ref{fig: FQH En}(b) \blue{(for details of the computation, see Secs.~\ref{sec: FQH L0}~and~\ref{sec: FQH L0+LJ}).}
The parameters are set to the same values as those of Fig.~\ref{fig: FQH En}.
For the computation of the Chern number, we have employed the method proposed in Ref.~\onlinecite{Fukui_FHS_mechod_JPSJ05}.
}
\label{fig: FQH Ch}
\end{figure}

We analyze the above bilayer system (\ref{eq: C-iso L nHFQH}) by introducing a parameter $\lambda$ ($0 \leq \lambda \leq 1$), $\matf{L}(\lambda):=\matf{L}_0+\lambda \matf{L}_{\mathrm{J}}$.
Employing the pseudo-potential approach (see Sec.~\ref{sec: FQH L0} and Appendix~\ref{sec: pp Liou app}), we obtain the spectrum and the pseudo-spin Chern number which are shown in Figs.~\ref{fig: FQH En}~and~\ref{fig: FQH Ch}.
As discussed in Sec.~\ref{sec: FQH L0+LJ}, these figures indicate that the topological properties of the non-Hermitian FQH states remain unchanged even in the presence of the jump term; the topological degeneracy and the pseudo-spin Chern number are not affected by the jump term.

Because the open quantum system loses but does not gain particles, \blue{the vacuum ($|\rho\rrangle =|0\rrangle_a\otimes |0 \rrangle_b$ with $|0\rrangle_\sigma$ being the state annihilated by all $d_{i\sigma}$) has an infinite lifetime}, which is consistent with Fig.~\ref{fig: FQH En}.
Namely, the Laughlin states, which are indicated by dots marked with the arrow, are no longer the states with the longest lifetime.
We note, however, that the topology of the Laughlin states is maintained even in the presence of the jump term.
Such topological states are considered to be experimentally accessible by observing the transient dynamics of cold atoms.
The realization of Laughlin states in cold atoms has been theoretically proposed~\cite{Sorensen_fluxRot_PRL05,Jaksch_SynthGauge_2003,Mueller_SynthGauge_PRA04}. 
Following these proposals, one can prepare the Laughlin state as the initial state for a sufficiently deep trap potential.
Suddenly making the trap potential shallower results in two-body loss. 
Furthermore, the non-Hermitian FQH states become the first decay modes by tuning the gauge field so that $N_\phi=\phi L_xL_y=6$ is satisfied.

As we see below, our numerical results demonstrate that both the spectral and the topological properties are encoded into the effective non-Hermitian Hamiltonian if $\matf{L}_\mathrm{J}$ and $\matf{L}_0$ are written in block-upper-triangular and block-diagonal forms, respectively. 
The analysis of $H_{\mathrm{eff}}$ is numerically less demanding than that of the full Liouvillian $\matf{L}=\matf{L}_0+\matf{L}_{\mathrm{J}}$.

\subsubsection{
Results in the absence of the jump term
}
\label{sec: FQH L0}

Firstly, we discuss the case of $\matf{L}(0)=\matf{L}_0$ which can be understood from the previous work~\cite{Yoshida_nHFQH19} for the effective non-Hermitian Hamiltonian $H_{\mathrm{eff}}$. 

Let $E_{n_1}$ ($n_1=1,2\cdots$) be eigenvalues of $H_{\mathrm{eff}}$.
Because the state with the minimum real-part of the energy $\mathrm{Re}E_{n_1}$ also shows the longest lifetime, $1/\mathrm{Im}E_{n_1}$, the pseudo-potential approach is employed where the creation operator $c^\dagger_i$ is replaced to $f^\dagger_i=\sum'_{n_1} \varphi^*_{in_1} a^\dagger_{n_1}$ (for more details, see Appendix~\ref{sec: pp Liou app}).
Here, $\varphi_{in_1}$ denotes a state in the lowest Landau level; $\sum_j h_{ij}\varphi_{jn_1}=\varphi_{in_1} \epsilon_{n_1}$ with the energy $\epsilon_{n_1}\in \mathbb{R}$. The operator $a^\dagger_{n_1}$ creates a fermion with a state in the lowest Landau level.
The summation $\sum'_{n_1}$ is taken over states in the lowest Landau level.
Diagonalizing $H_{\mathrm{eff}}$ for the filling factor $\nu=1/3$ for the lowest Landau level, we can observe the three-fold degeneracy for the states with the longest lifetime~\cite{Haldane_TopoDeg_PRL85,Yoshida_nHFQH19}, which is the topological degeneracy of the Laughlin states for $\nu=1/3$. 
We note that the number of fermions is conserved in the absence of the jump term.
For these three-fold degenerate states, the Chern number defined in Eq.~(\ref{eq: state Ch}) takes one ($C_{H_{\mathrm{eff}}}=1$)~\blue{\cite{Yoshida_nHFQH19,cond_mbdyCh_ftnt,Niu_HallCond_PRB85}, which indicates the robustness of the  topology against the non-Hermiticity.} 

With the above facts, we can understand the results of $\matf{L}_0$ which can be block-diagonalized into each subsector labeled by $(N_a,N_b)$ with $N_\sigma$ denoting the total number of fermions in layer $\sigma=a,b$.
In Fig.~\ref{fig: FQH En}, the colored dots represent the spectrum of $\matf{L}_0$ which is given by $\Lambda_n=E_{n_1}-E^*_{n_2}$ with $E_{n_{1(2)}}$ denoting the eigenvalues of $H_{\mathrm{eff}}$.
The states indicated by dots marked with the arrow correspond to the Laughlin states at the filling factor $\nu=1/3$.
Here, we note that these states show 9-fold degeneracy $(N^2_d=9)$ because there is topologically protected three-fold degeneracy $(N_d=3)$ for each of the two layers.
We also note that the data for $N_a=4$ is similar to those of $N_a=2$, which is attributed to the pseudo-potential approach projecting creation operators onto the states in the lowest Landau level~\cite{nu2ov3_nu1ov3_ftnt}.
Figure~\ref{fig: FQH Ch} shows that the pseudo-spin Chern number for these 9-fold degenerate states takes three at $\lambda=0$, which is consistent with $C_{H_{\mathrm{eff}}}=1$. Namely, $C_{\mathrm{ps}}=N_dC_{H_\mathrm{eff}}=3$ holds with $N_d=3$ [see Eq.~(\ref{eq: Ckk to state Ch})].

\subsubsection{
Results in the presence of the jump term
}
\label{sec: FQH L0+LJ}
Let us now analyze the case for a finite value of $\lambda$ ($0 < \lambda \leq 1$).
We show that: (i) topological degeneracy is maintained; (ii) the pseudo-spin Chern number remains one for the non-Hermitian FQH states.

The topological degeneracy ($9$-fold degeneracy) survives even in the presence of the jump term. 
This is because the spectrum is not affected by the jump term $\matf{L}_\mathrm{J}$ when $\matf{L}_\mathrm{J}$ and $\matf{L}_0$ can be written in block-upper-triangular and block-diagonal forms, respectively~\cite{Torres_Utrian_PRA14,Nakagawa_exactHubb_arXiv20} (see Appendix~\ref{sec: triangular app});
for the open quantum system with two-body loss but without gain, the jump term $\matf{L}_\mathrm{J}$ maps states in the subspace labeled by $(N_a+2,N_b+2)$ to those in subspaces labeled by $(N_a,N_b)$, while $\matf{L}_0$ is block-diagonalized for subspaces labeled by $(N_a,N_b)$.
The numerical data for two-body loss also support the above independence of the spectrum.
In Fig.~\ref{fig: FQH En}, we can see that the eigenvalues of $\matf{L}_0$ (colored dots) and those of $\matf{L}=\matf{L}_0+\matf{L}_{\mathrm{J}}$ (black dots) are exactly on top of each other.
We note that the spectrum of $\matf{L}$ is obtained for the subsector labeled by $N_a-N_b$ and $(-1)^{N_a}$ where the ``Liouvillian" $\matf{L}(\lambda)$ is block-diagonalized.
The above numerical data show that the topological degeneracy survives even in the presence of the jump term, which is expected on general grounds.

The pseudo-spin Chern number should not be affected by the jump term, as the gap-closing does not occur.
Indeed, Fig.~\ref{fig: FQH Ch} indicates that the pseudo-spin Chern number takes three for an arbitrary value of $\lambda$ ($0 \leq \lambda \leq 1$).
Noting the relation $C_{\mathrm{ps}}=3C_{H_\mathrm{eff}}$ [see Eq.~(\ref{eq: Ckk to state Ch})], we conclude that topological properties of $H_{\mathrm{eff}}$ \blue{remain} unchanged even in the presence of the jump terms. 
Figure~\ref{fig: FQH Ch} is obtained by employing the method proposed in Ref.~\onlinecite{Fukui_FHS_mechod_JPSJ05}.

In the above, we have confirmed that the topological properties of the Laughlin state are maintained even in the presence of the jump term. 
Furthermore, the above results elucidate that both the spectral and the topological properties are encoded into the effective non-Hermitian Hamiltonian if $\matf{L}_\mathrm{J}$ and $\matf{L}_0$ are written in block-upper-triangular and block-diagonal forms, respectively. 

We close this section with a remark on the topological degeneracy; for another type of Lindblad operators preserving the charge $\mathrm{U(1)}$ symmetry, e.g., the Lindblad operators describing dephasing noise~\cite{Cai_dephasing_PRL13,Znidaric_Choiiso_PRE15,Prosen_exactHubb_PRL16,Caspel_dephasing_PRA18,Shibata_Choiiso_PRB19}, three-fold topological degeneracy can be observed (for more details, see Appendix~\ref{sec: topo deg jump app}).

\section{Summary}
Despite the previous extensive analysis of open quantum systems, it is unclear whether correlated topological states, such as FQH states, are maintained even in the presence of the jump term.

In this paper, we have introduced the pseudo-spin Chern number computed from the vectorized density matrices in the doubled Hilbert space $\mathrm{Ket}\otimes\mathrm{Bra}$ which is akin to the spin-Chern number.
The presence of such a topological invariant elucidates that as long as the gap of ``Liouvillian" $\matf{L}(\lambda)=\matf{L}_{0}+\lambda \matf{L}_{\mathrm{J}}$ opens for $0 \leq \lambda \leq 1$, the topology of the full Liouvillian $\matf{L}(1)$ is encoded into $H_{\mathrm{eff}}$.
The effective Hamiltonian is particularly useful for systems where $\matf{L}_{\mathrm{J}}$ and $\matf{L}_{0}$ can be written in block-upper-triangular and block-diagonal forms, respectively. This is because in such systems, both the spectral and topological properties are encoded into the effective Hamiltonian.

As an application, we have addressed the topological characterization of the non-Hermitian FQH states in open quantum systems with two-body loss but without gain. 
Our numerical results have elucidated that even in the presence of the jump term, topological properties (i.e., the pseudo-spin Chern number and 9-fold topological degeneracy) of the non-Hermitian FQH states are not affected by the jump term.
This fact also reduces the numerical cost because the analysis of $H_{\mathrm{eff}}$ is numerically less demanding than that of the full Liouvillian $\matf{L}=\matf{L}_0+\matf{L}_{\mathrm{J}}$.

We note that similar topological invariants can be introduced to characterize correlated topological states for other spatial dimensions and symmetry [e.g., a one-dimensional open quantum systems with inversion symmetry (see Appendix~\ref{sec: 1D L Berry app})], indicating the high versatility of our approach.

\section*{
Acknowledgements
}
The authors thank Masaya Nakagawa for fruitful discussion. 
This work is supported by JSPS Grant-in-Aid for Scientific Research on Innovative Areas ``Discrete Geometric Analysis for Materials Design": Grants No.~JP20H04627 and No.~JP20H04630.
This work is also supported by JSPS KAKENHI Grants No.~JP16K13845, No.~JP17H06138, No~JP18K03445, No.~JP18H05842, No.~19K21032, and No.~JP19J12317.
A part of numerical calculations were performed on the supercomputer at the ISSP in the University of Tokyo.


\begin{thebibliography}{138}
\bibitem [{cla()}]{classA_ftnt}%
  \BibitemOpen
  \href@noop {} {}\bibinfo {note} {{ In the context of non-interacting
  topological insulators, it is known that systems without symmetry belong to
  class~A }}\BibitemShut {NoStop}%
\bibitem [{int()}]{interp_Liouvi_ftnt}%
  \BibitemOpen
  \href@noop {} {}\bibinfo {note} {{ \blue{ We note that the interpolation is
  just to examine the topological equivalence between the states of
  $\spopf{L}(\lambda=0)$ and $\spopf{L}(\lambda=1)$; the ``Liouvillian"
  $\spopf{L}(\lambda)$ describes the open quantum dynamics only for
  $\lambda=0,1$ } }}\BibitemShut {NoStop}%
\bibitem [{gap()}]{gap_ftnt}%
  \BibitemOpen
  \href@noop {} {}\bibinfo {note} {{ The pseudo-spin Chern number defined below
  can change when $\Lambda_n = \Lambda_{n'}$ is satisfied. However, we define
  the gap as Eq.~(\ref{eq: def of gap}) by focusing on the imaginary-part which
  is related to the lifetime of the states in open quantum systems
  }}\BibitemShut {NoStop}%
\bibitem [{top()}]{topo_deg_defs_ftnt}%
  \BibitemOpen
  \href@noop {} {}\bibinfo {note} {{ \blue{ By topological ordered states, we
  mean the states that exhibit topological degeneracy depending on the genus of
  the system. Typical examples are FQH states } }}\BibitemShut {NoStop}%
\bibitem [{inv()}]{inviso_ftnt}%
  \BibitemOpen
  \href@noop {} {}\bibinfo {note} {{ We note that this map may have ambiguity
  of the phase factor. However, it does not affect the definition of the
  pseudo-spin Chern number which is gauge invariant }}\BibitemShut {NoStop}%
\bibitem [{cba()}]{cbar_ftnt}%
  \BibitemOpen
  \href@noop {} {}\bibinfo {note} {{ Noting that $ \bar{c}^\dagger_{i'}
  c_{i}|\phi_{j_1} \rrangle_{K} \otimes|\phi_{j_2}\rrangle_{B} =
  c_{i}|\phi_{j_1} \rrangle_{K} \otimes \left(
  \bar{c}^\dagger_{i}|\phi_{j_2}\rrangle_{B} \right) $ holds, we have $
  {}[c_i,\bar{c}^\dagger_{i'}]=0 $ }}\BibitemShut {NoStop}%
\bibitem [{Bos()}]{Bosonization_ftnt}%
  \BibitemOpen
  \href@noop {} {}\bibinfo {note} {{ A similar procedure is taken in
  Ref.~\onlinecite{Freericks_PRB95} where fermions are mapped to bosons. We
  note that the introduction of $d_{ib}$ is innocuous, provided that $P_{fa}$
  commutes with $\matf{L}=\matf{L}_0+\matf{L}_{\mathrm{J}}$; $[ P_{fa},
  \matf{L}]=0$ }}\BibitemShut {NoStop}%
\bibitem [{con()}]{cond_mbdyCh_ftnt}%
  \BibitemOpen
  \href@noop {} {}\bibinfo {note} {{ \blue{ In the Hermitian case, it is
  well-known that for the three-fold degenerate FQH states, the many-body Chern
  number taking one is a source of the fractional Hall conductance
  $\sigma_{xy}=1/3$ (see Ref.~\onlinecite{Niu_HallCond_PRB85}) } }}\BibitemShut
  {NoStop}%
\bibitem [{nu2()}]{nu2ov3_nu1ov3_ftnt}%
  \BibitemOpen
  \href@noop {} {}\bibinfo {note} {{ With this approximation, we can see that
  the particle-hole transformation maps the Hamiltonian $H_{\mathrm{eff}}$ for
  $\nu=2/3$ to that for $\nu=1/3$. Thus, supposing that $E_{n_1}$ denotes the
  energy eigenvalue for $\nu=1/3$, the energy eigenvalue for $\nu=2/3$ can be
  written as $E_{n_1}+E'$ with a complex number $E'$ }}\BibitemShut {NoStop}%
\bibitem [{Ber()}]{Berry_SSH_dephasing_ftnt}%
  \BibitemOpen
  \href@noop {} {}\bibinfo {note} {{ By using Eq.~(\ref{eq: SSH rhog_theta
  PBC}), we can compute the integral as follows: $ \chi_{\uparrow g} =
  \int^{\pi}_{-\pi} \!\! d\theta \, \mathrm{Im} {}_{L}\llangle \rho_g (\theta)|
  \partial^\uparrow_\theta |\rho_g(\theta)\rrangle_{R} =\int^{\pi}_{-\pi} \!\!
  d\theta \, \llangle -2| n_{1A\uparrow} |-2\rrangle = \pi $ }}\BibitemShut
  {NoStop}%
\bibitem [{upt()}]{uptri_ftnt}%
  \BibitemOpen
  \href@noop {} {}\bibinfo {note} {{ Here, we have used the following relation:
  $ \mathrm{det} \left( \begin{array}{cc} A & C \\ 0 & B \end{array} \right) =
  \mathrm{det}A\mathrm{det}B, $ where $A$ ($B$) is a $N \times N$ ($M\times M$)
  matrix, respectively. The matrix $C$ is a $N\times M$ }}\BibitemShut
  {NoStop}%
\bibitem [{def()}]{defs_adag_a_ftnt}%
  \BibitemOpen
  \href@noop {} {}\bibinfo {note} {{ The operators $a^\dagger_{n_{l}k_{l'}a}$
  and $a^\dagger_{n_{l}k_{l'}b}$ are defined as
  $a^\dagger_{n_{l}k_{l'}a}:=a^\dagger_{n_{l}k_{l'}}$ and
  $a^\dagger_{n_lk_{l'}b}:=\bar{a}^\dagger_{n_{l}k_{l'}}(-1)^{N_a}$ with
  $N_a=\sum_{n_{l}k_{l'}}a^\dagger_{n_{l}k_{l'}a}a_{n_{l}k_{l'}a}$. Here,
  $a^\dagger_{n_{{l}}k_{l'}}$ creates a fermion in state $|\varphi_{n_{l}
  }(k_{yl'})\rangle$ which acts on the original Hilbert space (i.e.,
  $\mathrm{Ket}$ space). The operator $\bar{a}^\dagger_{n_{l}k_{l'}}$ acts on
  the subspace $\mathrm{Bra}$; for an arbitrary vector $|\rho\rrangle$,
  $\bar{a}^\dagger_{n_{l}k_{l'}} |\rho\rrangle $ is identified as $ \rho
  a_{n_{l}k_{l'}}$ }}\BibitemShut {NoStop}%
\end{thebibliography}
%


\appendix

\section{
Details of the isomorphism defined in Eq.~(\ref{eq: choi iso rho})
}
\label{sec: Choi iso app}
With the isomorphism [see Eq.~(\ref{eq: choi iso rho})], the action of the Liouvillian $\spopf{L}[\, \cdot \, ]$ on a density matrix is mapped to a vector as follows: 
\begin{subequations}
\label{eq: Liouv Choi app}
\begin{eqnarray}
\spopf{L}[\rho(t)] & \leftrightarrow & \matf{L} |\rho(t)\rrangle,
\end{eqnarray}
with
\begin{eqnarray}
\matf{L} &=& \matf{L}_0+\matf{L}_{\mathrm{J}}, \\
\matf{L}_0 &=& \left( H_{\mathrm{eff}}\otimes \1- \1 \otimes H^*_{\mathrm{eff}} \right), \\
\matf{L}_{\mathrm{J}} &=& i\sum_\alpha L_\alpha \otimes L^*_\alpha.
\end{eqnarray}
\end{subequations}
Here $\1$ denotes the identity operator. 

To see this, we first note that the isomorphism [see Eq.~(\ref{eq: choi iso rho})] maps the density matrix $\rho \in \mathrm{End}_\mathbb{C}(\mathcal{H})$, which act on the Hilbert space $\mathcal{H}$, to the vector in the doubled Hilbert space $|\rho \rrangle \in \mathrm{Ket} \otimes \mathrm{Bra}$. 
Correspondingly, the superoperator $\spopf{L} \in \mathrm{End}_{\mathbb{C}}\left(\mathrm{End}_{\mathbb{C}}(\mathcal{H})\right)$ is mapped to a non-Hermitian matrix $\matf{L}$. 
In particular, we have
\begin{eqnarray}
&&A \rho B
=\sum_{iji'j'} A_{i'i}\rho_{ij}B_{jj'} |\phi_{i'}\rangle \langle \phi_{j'}| \nonumber \\
&&\leftrightarrow 
\sum_{iji'j'} (A_{i'i}\otimes B^T_{j'j}) \rho_{ij} |\phi_{i'}\rrangle_{K} \otimes |\phi_{j'}\rrangle_{B}=A\otimes B^T |\rho\rrangle, \nonumber\\ 
\end{eqnarray}
where $A_{ij}:=\langle \phi_i | A |\phi_j \rangle$, $B_{ij}:=\langle \phi_i | B |\phi_j \rangle$ with $|\phi_j \rangle$ being the set of states generated by acting on the vacuum with creation operators in the real space
[e.g., for spinless fermions, $|\phi_i\rangle$ is generated by acting with the creation operators $c^\dagger_{j}$ ($j=1,2,\cdots$) on the vacuum].
By making use of the above relation, we have 
\begin{subequations}
\begin{eqnarray}
 \rho H^\dagger_{\mathrm{eff}} &\leftrightarrow&  \1 \otimes (H^\dagger_{\mathrm{eff}})^T |\rho \rrangle, \\
 L_\alpha \rho L^\dagger_{\alpha} &\leftrightarrow& L_\alpha \otimes (L^\dagger_\alpha)^T |\rho \rrangle.
\end{eqnarray}
\end{subequations}
Therefore, we can see that the Liouvillian $\spopf{L}[\rho(t)]$ is mapped to a non-Hermitian matrix $\matf{L}$ as shown in Eq.~(\ref{eq: Liouv Choi app}).

\section{
Characterization of one-dimensional open quantum systems with inversion symmetry
}
\label{sec: 1D L Berry app}
In Sec.~\ref{sec: ps Chern}, we have introduced the pseudo-spin Chern number to characterize topological properties maintained even in the presence of the jump term for two-dimensional open quantum systems without symmetry.
The pseudo-spin Chern number can be computed by twisting the boundary condition either $\mathrm{Ket}$ or $\mathrm{Bra}$ space.
We show that this approach can be straightforwardly applied to one-dimensional open quantum systems with inversion symmetry, in which case the Berry phase is quantized to $0$ or $\pi$.
The presence of such a quantized topological invariant elucidates that the topology of the full Liouvillian is encoded into $H_{\mathrm{eff}}$ when $\matf{L}_{\mathrm{J}}$ and $\matf{L}_0$ can be written in block-upper-triangular and block-diagonal forms, respectively.
This fact is particularly useful for systems with loss but without gain as demonstrated in Sec.~\ref{sec: App nHFQH}.

As an application to one-dimensional open quantum systems with dissipation, we analyze the Su-Schrieffer-Heeger (SSH) model with dephasing noise whose topology has not been characterized so far.

\subsection{
Berry phase for open quantum systems
}
\label{sec: 1D berry phase app}

\subsubsection{
Definition
}
\label{sec: def Berry phase}

Let $\matf{L}(\theta)$ be a one-parameter family of Liouvillian depending smoothly on $\theta$ and periodic in $\theta$, i.e., $\matf{L}(\theta+2\pi)= \matf{L}(\theta)$.
Here, $\theta$ dependence is introduced only for the subspace $\mathrm{Ket}$.
We assume that there exists a $\theta$-independent operator $I$ such that $I^2=1$ and $I \matf{L}(\theta) I = \matf{L}(-\theta)$.
The Berry phase introduced in this section is available regardless whether the particles are fermions or bosons.

Suppose that the right and left vectors of the Liouvillian, $| \rho_n (\theta) \rrangle_R$ and ${}_L \llangle \rho_n(\theta) |$, are non-degenerate.
In this case, choosing the gauge so that $| \rho_n (\theta+2\pi) \rrangle_R=| \rho_n (\theta) \rrangle_R$ and ${}_L \llangle \rho_n(\theta+2\pi) |={}_L \llangle \rho_n(\theta) |$ are satisfied, we can define the following Berry phase
\begin{subequations}
\label{eq: def of chi_k app}
\begin{eqnarray}
\chi_{Kn} &=& \int^{\pi}_{-\pi} \!\!  d\theta \ \mathrm{Im} A_{Kn}(\theta), \\
A_{Kn}(\theta)  &=& {}_L \llangle \rho_n(\theta) | \partial^{K}_\theta | \rho_n(\theta) \rrangle_R.
\end{eqnarray}
\end{subequations}
Here $\partial^{K}_\theta$ denotes the derivative with respect to $\theta$ which acts only on the subspace $\mathrm{Ket}$; for instance, the action of $\partial^K_\theta$ on a state $|\Phi\rrangle_K\otimes |\Psi'\rrangle_B$ reads $(\partial^K_\theta|\Phi\rrangle_K)\otimes |\Psi'\rrangle_B$.
We have imposed the biorthogonal normalization condition on the right and left eigenvectors of $\matf{L}(\theta)$; $|\rho_n (\theta) \rrangle_R$ and ${}_L\llangle \rho_{n'}(\theta)  |$ satisfy ${}_L \llangle \rho_{n'}(\theta) |\rho_n(\theta) \rrangle_R=\delta_{n'n}$ for arbitrary integers, $n$ and $n'$.
%

\subsubsection{
Properties of the Berry phase $\chi_{Kn}$
}
\label{sec: properties Berry phase}

The Berry phase $\chi_{Kn}$ elucidates that as long as the gap of the ``Liouvillian" $\matf{L}(\lambda)$ opens for $0 \leq \lambda \leq 1$, the topological properties of $H_{\mathrm{eff}}$ are maintained even in the presence of the jump term, which follows from the following two facts.

(i) The Berry phase is quantized,
\begin{eqnarray}
\label{eq: chi Z2 app}
e^{i\chi_{Kn}} &=& \prod_{\theta_0=0,\pi} {}_L\llangle \rho_n(\theta_0) |I|\rho_n(\theta_0)\rrangle_R \in \{ -1, 1\},
\end{eqnarray}
where the right eigenvector $|\rho_n(\theta_0)\rrangle_R$ is also a right eigenvector of $I$ with an eigenvalue $\pm1$ for $\theta_0=0$ or $\pi$.
Equation~(\ref{eq: chi Z2 app}) is proven in Appendix~\ref{sec: proof Berry phase app}.

(ii) In the absence of the jump term, $\chi_{Kn}$ is written as
\begin{eqnarray}
\label{eq: chi Heff app}
\chi_{Kn}&=& \int^\pi_{-\pi} \!\! d\theta \ \mathrm{Im} {}_L \langle \Phi_{n_1} | \frac{\partial}{\partial \theta} | \Phi_{n_1} \rangle_R,
\end{eqnarray}
where $|\Phi_{n_1} \rangle_R$ and ${}_L \langle \Phi_{n_1} |$ ($n_1=1,2,\cdots$) are the right and left eigenvectors of $H_{\mathrm{eff}}(\theta)$,
\begin{subequations}
\label{eq: Berry Heff |R> =En |R> app}
\begin{eqnarray}
H_{\mathrm{eff}}(\theta) | \Phi_{n_1}(\theta) \rangle_R &=& E_{n_1}(\theta)| \Phi_{n_1}(\theta) \rangle_R, \\
{}_L \langle \Phi_{n_1}(\theta) | H_{\mathrm{eff}}(\theta) &=& {}_L \langle \Phi_{n_1}(\theta) | E_{n_1}(\theta),
\end{eqnarray}
\end{subequations}
with the eigenvalue $E_{n_1}(\theta) \in \mathbb{C}$.
Equation~(\ref{eq: Berry Heff |R> =En |R> app}) is proven in Appendix~\ref{sec: proof Berry phase app}.

Equation~(\ref{eq: chi Heff app}) indicates that $\chi_{Kn}$ is reduced to the Berry phase for $H_{\mathrm{eff}}$ in the absence of the jump term. 
In addition, Eq.~(\ref{eq: chi Z2 app}) indicates that as long as the gap opens, $\chi_{Kn}$ does not change its value even when the jump term is introduced.
Therefore, the Berry phase $\chi_{Kn}$ elucidates that as long as the gap of the ``Liouvillian" $\matf{L}(\lambda)$ opens for $0 \leq \lambda \leq 1$, the topological properties of $H_{\mathrm{eff}}$ are maintained even in the presence of the jump term. 

In particular, this fact indicates that the topology of the full Liouvillian is encoded into $H_{\mathrm{eff}}$ when $\matf{L}_{\mathrm{J}}$ and $\matf{L}_0$ can be written in block-upper-triangular and block-diagonal forms, respectively. 
An example of such systems is an open quantum system with loss but without gain, as we have seen in Sec.~\ref{sec: App nHFQH} where the two-dimensional system is analyzed.

We note that Berry phases for non-Hermitian systems are defined in several contexts~\cite{Lieu_BerryPT_PRB18,Dangel_BerryPT_PRA18}.
However, it remained unsolved whether there exists a topological invariant that characterizes the topological properties even in the presence of the jump term.

In the rest of this section, we prove Eqs.~(\ref{eq: chi Z2 app})~and~(\ref{eq: chi Heff app}).

\subsubsection{
Proof of Eqs.~(\ref{eq: chi Z2 app})~and~(\ref{eq: chi Heff app})
}
\label{sec: proof Berry phase app}

\textit{
Proof of Eq.~(\ref{eq: chi Z2 app})--.
}
For the inversion symmetric system satisfying $I \matf{L}(\theta) I^{-1} = \matf{L}(-\theta)$, the following relation holds:
\begin{subequations}
\label{eq: I |R> and <L|I app}
\begin{eqnarray}
I | \rho_n  (-\theta) \rrangle_R &=&  | \rho_n  (\theta) \rrangle_R c_{n}(\theta),\\
{}_L \llangle \rho_n (-\theta)| I &=&  c^{-1}_{n}(\theta) {}_L \llangle \rho_n (\theta)|,
\end{eqnarray}
\end{subequations}
with a continuous function $c_{n}(\theta)$ taking a complex value $c_{n}(\theta) \neq 0$.
We recall the assumption that the right and left eigenvectors are non-degenerate.
By using the above relation, we can obtain
\begin{eqnarray}
A_{Kn}(-\theta)  &=& {}_L \llangle \rho_n(-\theta) | \partial^{K}_{-\theta}  | \rho_n(-\theta) \rrangle_R \nonumber \\
              &=& -{}_L \llangle \rho_n(-\theta) | \partial^{K}_\theta | \rho_n(-\theta) \rrangle_R \nonumber \\
              &=& -{}_L \llangle \rho_n(-\theta) | I \partial^{K}_\theta I | \rho_n(-\theta) \rrangle_R \nonumber \\
              &=& -c^{-1}_{n}(\theta){}_L \llangle \rho_n(\theta) | \partial^{K}_\theta | \rho_n(\theta) \rrangle_R c_{n}(\theta) \nonumber \\
              &=& -A_{Kn}(\theta) - c^{-1}_{n}(\theta)\frac{\partial}{\partial \theta } c_{n}(\theta).   
\end{eqnarray}
This relation simplifies the integral in Eq.~(\ref{eq: def of chi_k app}a),
\begin{eqnarray}
\label{eq: chi_k = c(pi) -c(0) app}
\chi_{Kn} &=& \int^{0}_{-\pi}  \!\!\! d\theta \ \mathrm{Im} A_{Kn}(\theta) + \int^{\pi}_{0} \!\! d\theta \ \mathrm{Im} A_{Kn}(\theta) \nonumber \\
       &=& \int^{\pi}_{0}  \!\!\! d\theta \ \mathrm{Im} \left[ A_{Kn}(-\theta) + A_{Kn}(\theta) \right] \nonumber \\
       &=& -\int^{\pi}_{0}  \!\!\! d\theta \ \mathrm{Im}  c^{-1}_{n}(\theta) \frac{\partial}{\partial \theta} c_{n}(\theta)  \nonumber \\
       &=& -\mathrm{Im} \left[ \log c_{n}(\pi) -\log c_{n}(0) \right].
\end{eqnarray}
Equation~(\ref{eq: I |R> and <L|I app}) indicates that $| \rho_n  (0) \rrangle_R$ [$| \rho_n  (\pi) \rrangle_R$] is a right eigenvector of $I$ with eigenvalue $c_{n}(0)$ [$c_{n}(\pi)$]. 
Namely, $c_{n}(0)$ and $c_{n}(\pi)$ take $1$ or $-1$.
Therefore, combining this fact and Eq.~(\ref{eq: chi_k = c(pi) -c(0) app}), we obtain Eq.~(\ref{eq: chi Z2 app}) which indicates the quantization of the Berry phase $\chi_{Kn}$.

\textit{
Proof of Eq.~(\ref{eq: chi Heff app})--.
}
In the absence of the jump term, we can see the following correspondence
\begin{eqnarray}
 |\rho_n \rrangle_R \leftrightarrow   | \Phi_{n_1} \rangle_R {}_R\langle \Phi_{n_1} |, \quad {}_L \llangle \rho_n | \leftrightarrow  | \Phi_{n_1} \rangle_L {}_L\langle \Phi_{n_1} |.\nonumber \\
\end{eqnarray}
Here, we recall the assumption that the states are non-degenerate. 
By using the above correspondence, $\chi_{Kn}$ is written as
\begin{eqnarray}
\label{eq: proof of chi Heff app}
\chi_{Kn} &=& \int^{\pi}_{-\pi} \!\!\! d\theta \ \mathrm{Im} \mathrm{tr}\left( | \Phi_{n_1} \rangle_L {}_L\langle \Phi_{n_1} | \partial^{K}_\theta | \Phi_{n_1} \rangle_R {}_R\langle \Phi_{n_1} |     \right)  \nonumber \\
       &=& \int^{\pi}_{-\pi} \!\!\! d\theta \ \mathrm{Im} {}_L\langle \Phi_{n_1} | \frac{\partial}{\partial \theta} | \Phi_{n_1} \rangle_R,
\end{eqnarray}
which is the desired Eq.~(\ref{eq: chi Heff app}).

\subsection{
SSH model with dephasing noise
}
\label{sec: SSH app}
In the above, we have introduced the Berry phase for the doubled Hilbert space [see Eq.~(\ref{eq: def of chi_k app})]. 
In particular, the Berry phase elucidates that both the spectral and topological properties of the Liouvillian are encoded into the effective non-Hermitian Hamiltonian $H_{\mathrm{eff}}$ for open quantum systems whose jump term can be written in a block-upper-triangular form. 
This is because such a jump term does not affect the spectrum.

In this section, instead of the detailed analysis of such an open quantum system, we address topological characterization of a one-dimensional system with dephasing noise~\cite{Cai_dephasing_PRL13,Znidaric_Choiiso_PRE15,Prosen_exactHubb_PRL16,Caspel_dephasing_PRA18,Shibata_Choiiso_PRB19}, demonstrating that our topological invariant works even when the jump term affects the spectrum of the Liouvillian.
Specifically, we analyze the SSH model with dephasing noise whose topological properties have not been analyzed so far.
Our analysis elucidates that a non-equilibrium steady state is characterized by the Berry phase taking $\pi$ in the presence of the jump term although the gap is closed in the absence of the jump term.

\subsubsection{
Mapping the open quantum system to a closed system
}

Consider the SSH model with dephasing noise described by the Lindblad equation~(\ref{eq: Lindblad}) with
\begin{subequations}
\begin{eqnarray}
\!\!\!\!\!\!\!\!\!\!\!\! H_0 &=& \sum^{L-1}_{j=0} tc^\dagger_{j+1A} c_{jB} + t' c^\dagger_{jA} c_{jB} +h.c., \\
\!\!\!\!\!\!\!\!\!\!\!\! L_{j\alpha} &=& \frac{\sqrt{\gamma}}{2}( c^\dagger_{j\alpha} c_{j\alpha}-c_{j\alpha}c^\dagger_{j\alpha})=\sqrt{\gamma}(c^\dagger_{j\alpha}c_{j\alpha}-\frac{1}{2}).
\end{eqnarray}
\end{subequations}
Here, $c^\dagger_{j\alpha}$ ($c_{j\alpha}$) creates (annihilates) a spinless fermion at sublattice $\alpha=A,B$ of site $j$.
Hopping integrals $t$ and $t'$ take real values, and  $\gamma$ is a positive number.
The number of unit cells is $L$.
We have imposed the periodic boundary condition $c^\dagger_{L\alpha}= c^\dagger_{0\alpha}$.

The above open quantum system is mapped to the closed system which has been discussed for the specific choice of $t'$ ($t'=t$)~\cite{Prosen_exactHubb_PRL16,Ziolkowska_exactHubb_arXiv19}.
The Liouvillian reads
\begin{subequations}
\label{eq: Liou 1D Hubb app}
\begin{eqnarray}
\matf{L}&=& \matf{L}_0+\matf{L}_{\mathrm{J}}, \\
\matf{L}_0 &=& \sum_{\alpha\sigma} \sum^{L-1}_{j=0}( t d^\dagger_{j+1A\sigma} d_{jB\sigma}+t' d^\dagger_{jA\sigma} d_{jB\sigma}   \nonumber \\
&& \qqqqquad \qqqqquad  +h.c.) -\frac{i\gamma L}{2},  \\
\matf{L}_{\mathrm{J}}&=& -i\gamma \sum_{\alpha}\sum^{L-1}_{j=0} (n_{j\alpha\uparrow}-\frac{1}{2}) (n_{j\alpha\downarrow}-\frac{1}{2}).
\end{eqnarray}
\end{subequations}
Here, we have used $\sigma=\uparrow$ ($\sigma=\downarrow$) to specify the subspace $\mathrm{Ket}$ ($\mathrm{Bra}$).
We denote by $d^\dagger_{j\alpha\sigma}$ the creation operator of a fermion with spin $\sigma=\uparrow,\downarrow$ at sublattice $\alpha=A,B$ of site $j$.
The number operator is defined as $n_{j\alpha\sigma}:=d^\dagger_{j\alpha\sigma}d_{j\alpha\sigma}$.

Now, we derive Eq.~(\ref{eq: Liou 1D Hubb app}). With the isomorphism [see Eq.~(\ref{eq: choi iso rho})], the following relations hold for an arbitrary density matrix $\rho$
\begin{subequations}
\begin{eqnarray}
 && \rho  \left(c^\dagger_{i\alpha}c_{i\alpha}-\frac{1}{2}\right)\left(c^\dagger_{i\alpha}c_{i\alpha}-\frac{1}{2}\right) \nonumber \\
 && \quad \quad \leftrightarrow \left(\bar{c}^\dagger_{i\alpha}\bar{c}_{i\alpha}-\frac{1}{2}\right)\left(\bar{c}^\dagger_{i\alpha}\bar{c}_{i\alpha}-\frac{1}{2}\right) |\rho \rrangle, \\
 && \left(c^\dagger_{i\alpha}c_{i\alpha}-\frac{1}{2}\right) \rho \left(c^\dagger_{i\alpha}c_{i\alpha}-\frac{1}{2}\right)  \nonumber \\
 && \quad \quad \leftrightarrow \left(c^\dagger_{i\alpha}c_{i\alpha}-\frac{1}{2}\right) \left(\bar{c}^\dagger_{i\alpha}\bar{c}_{i\alpha}-\frac{1}{2}\right) |\rho \rrangle,
\end{eqnarray}
\end{subequations}
where $c_{i\alpha}$ ($\bar{c}_{i\alpha}$) acts on the vectors in the subspace $\mathrm{Ket}$ ($\mathrm{Bra}$).
Thus, introducing the following operators, 
\begin{subequations}
\begin{eqnarray}
d_{i\alpha \uparrow} &=& c_{i\alpha}, \\
d_{i\alpha \downarrow} &=& \bar{c}_{i\alpha}(-1)^{\sum_{i\alpha} d^\dagger_{i\alpha\uparrow}d_{i\alpha\uparrow}},
\end{eqnarray}
\end{subequations}
the Liouvillian can be written as
\begin{subequations}
\begin{eqnarray}
\matf{L}&=& \matf{L}_0+\matf{L}_{\mathrm{J}},  \\
\matf{L}_0 &=& \sum_{\alpha\sigma} \sum^{L-1}_{j=0} \mathrm{sgn}(\sigma) (t d^\dagger_{j+1A\sigma} d_{jB\sigma}+t' d^\dagger_{jA\sigma} d_{jB\sigma}  \nonumber \\
 && \tenquad  +h.c.) -\frac{i\gamma L }{2}, \nonumber \\
\matf{L}_{\mathrm{J}} &=& i\gamma \sum_{j\alpha} \left(n_{j\alpha\uparrow}-\frac{1}{2}\right) \left(n_{j\alpha\downarrow}-\frac{1}{2}\right),
\end{eqnarray}
\end{subequations}
with $\mathrm{sgn}(\sigma)$ taking $1$ ($-1$) for $\sigma=\uparrow$ ($\downarrow$).

Further applying the particle-hole transformation only for down-spin states, 
\begin{eqnarray}
\label{eq: particle-hole SSH app}
d^\dagger_{i\alpha\downarrow}\to d_{i\alpha\downarrow},
\end{eqnarray}
we end up with Eq.~(\ref{eq: Liou 1D Hubb app}).

Here, we define the Liouvillian $\matf{L}(\theta)$ for the SSH model which is necessary to compute the Berry phase. 
Twisting the hopping between sites $j=0$ and $j=1$ only for the subspace specified with $\sigma=\uparrow$, the Liouvillian $\matf{L}(\theta)$ is written as 
\begin{subequations}
\label{eq: L theta SSH gen app}
\begin{eqnarray}
\matf{L}(\theta)   &=& \matf{L}_0(\theta)+\matf{L}_{\mathrm{J}},\\
\matf{L}_0(\theta) &=& \sum_{\alpha\sigma}\left( \sum^{L-1}_{j=1} t d^\dagger_{j+1A\sigma} d_{jB\sigma}+ te^{i\theta_{\sigma}} d^\dagger_{1A\sigma} d_{0B\sigma} +h.c.\right) \nonumber \\
&&+t' \sum_{j\alpha\sigma}\left(d^\dagger_{jA\sigma} d_{jB\sigma} +h.c.\right) - \frac{i\gamma L }{2},
\end{eqnarray}
\end{subequations}
with $\theta_\sigma=\theta[1+\mathrm{sgn}(\sigma)]/2$.

\subsubsection{
Results for $t'=0$
}

By analyzing a simple case for $t'=0$, we show that in the bulk, the non-equilibrium steady state (i.e., the states with an infinite lifetime) is characterized by the Berry phase $\pi$. 
Correspondingly, for the open boundary condition, edge states result in the charge polarization only at edges.
We note that the gap is closed in the absence of the jump term.

\textit{(i) Bulk properties--.}
Let us consider the ``Liouvillian" $\matf{L}(\lambda)=\matf{L}_0+\lambda \matf{L}_{\mathrm{J}}$ under the periodic boundary condition.
Here, $\matf{L}_0$ and $\matf{L}_{\mathrm{J}}$ are defined in Eq.~(\ref{eq: Liou 1D Hubb app}).
This model preserves the total number of particles for each spin.

For $t'=0$, the problem is reduced to a two-site Hubbard model with the pure-imaginary interaction,
\begin{eqnarray}
\matf{L}_{2\mathrm{site}}(\lambda) &=& t \sum_\sigma d^\dagger_{1A\sigma} d_{0B\sigma} +h.c. \nonumber \\
 && - i \lambda \gamma \left[\left(n_{1A\uparrow}-\frac{1}{2}\right)\left(n_{1A\downarrow}-\frac{1}{2}\right) \right. \nonumber \\
 && \quad \left. +\left(n_{0B\uparrow}-\frac{1}{2}\right) \left(n_{0B\downarrow}-\frac{1}{2}\right) \right] -\frac{i\gamma}{2}. \nonumber \\
\end{eqnarray}

Here, let us focus on the half-filled case where the dynamics can be understood by diagonalizing $\matf{L}_{2\mathrm{site}}(\lambda)$ for the subsector labeled by $(N_\uparrow, N_\downarrow)=(1,1)$ with $N_\sigma:=n_{1A\sigma}+n_{0B\sigma}$.

Firstly, we define the basis
\begin{eqnarray}
\left \{ |+1\rrangle, |+2\rrangle, |-1\rrangle, |-2\rrangle \right\},
\end{eqnarray}
spanning the subspace labeled by $(N_\uparrow, N_\downarrow)=(1,1)$. 
Here, $|\pm 1\rrangle$ and $|\pm 2\rrangle$ are defined as
\begin{subequations}
\begin{eqnarray}
|\pm 1\rrangle &:=& \frac{1}{\sqrt{2}} \left(d^\dagger_{1A\uparrow}d^\dagger_{1A\downarrow} \pm d^\dagger_{0B\uparrow}d^\dagger_{0B\downarrow} \right) | 0 \rrangle, \\ 
|\pm 2\rrangle &:=& \frac{1}{\sqrt{2}} \left(d^\dagger_{1A\uparrow}d^\dagger_{0B\downarrow} \pm d^\dagger_{0B\uparrow}d^\dagger_{1A\downarrow} \right) | 0 \rrangle, 
\end{eqnarray}
\end{subequations}
with the vacuum $| 0 \rrangle$ satisfying $d_{1A \sigma}|0\rrangle=0$ and $d_{0B \sigma}|0\rrangle=0$ for $\sigma=\uparrow,\downarrow$.

In this basis, $\matf{L}_{2\mathrm{site}}(\lambda)$ is represented as
\begin{subequations}
\label{eq: 2site L matrix app}
\begin{eqnarray}
\!\!\!\!\! \matf{L}_{2\mathrm{site}}(\lambda)&=& 
\left(
\begin{array}{cc}
\matf{L}_+(\lambda) & 0  \\
0 & \matf{L}_-(\lambda)
\end{array}
\right),
 \\
\!\!\!\!\!\matf{L}_+(\lambda)&=& 
\left(
\begin{array}{cc}
-i\gamma (1+\lambda)/2 & 2t \\
2t & -i\gamma (1-\lambda)/2
\end{array}
\right),
 \\
\!\!\!\!\!\matf{L}_-(\lambda)&=& 
\left(
\begin{array}{cc}
-i\gamma (1+\lambda)/2 & 0 \\
0 & -i\gamma (1-\lambda)/2
\end{array}
\right).
\end{eqnarray}
\end{subequations}

Diagonalizing the matrix $\matf{L}_{2\mathrm{site}}(\lambda)$, we can see that the eigenvalues are written as
\begin{subequations}
\begin{eqnarray}
\Lambda_{+a}&=& [-i\gamma +\sqrt{ 16t^2-\lambda^2 \gamma^2 }]/2, \\
\Lambda_{+b}&=& [-i\gamma -\sqrt{ 16t^2-\lambda^2 \gamma^2 }]/2, \\
\Lambda_{-a}&=& -i\gamma(\lambda+1)/2, \\
\Lambda_{-b}&=& i\gamma(\lambda-1)/2.
\end{eqnarray}
\end{subequations}

\begin{figure}[!h]
\begin{minipage}{0.475\hsize}
\begin{center}
\includegraphics[width=1\hsize,clip]{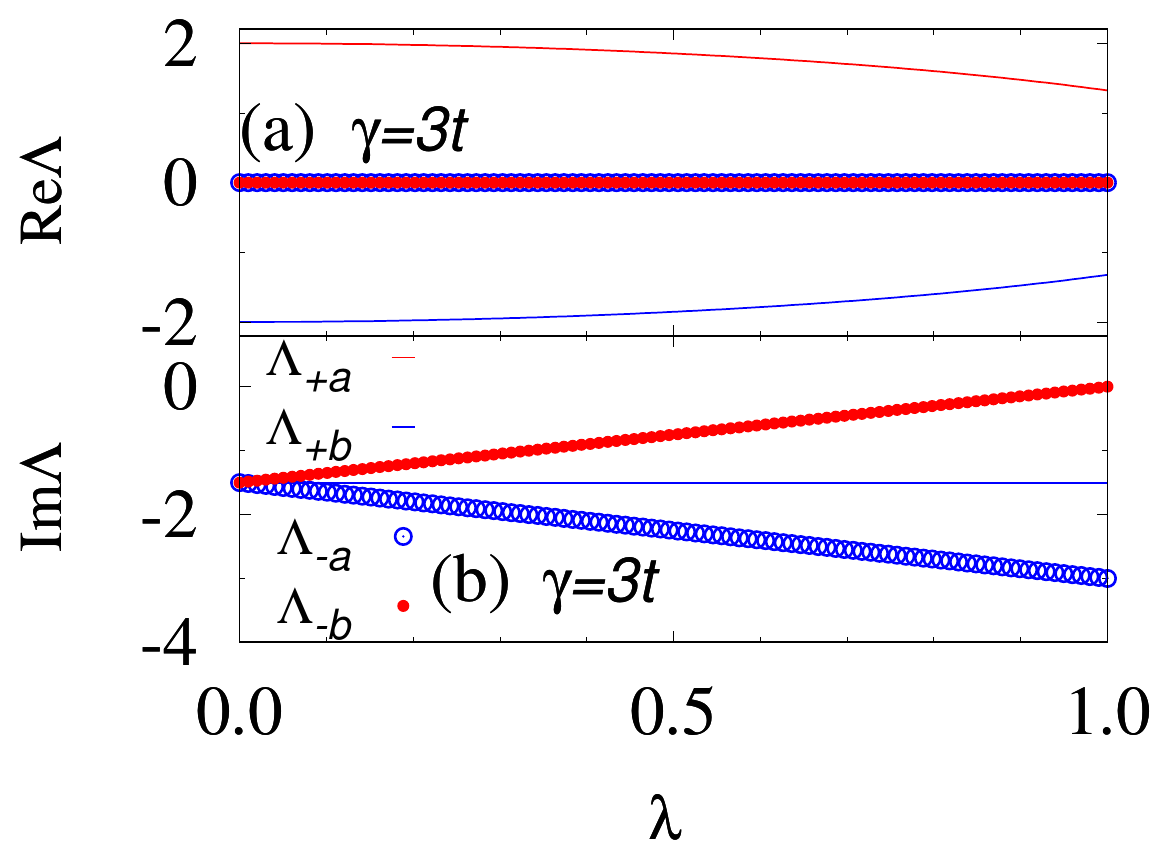}
\end{center}
\end{minipage}
\begin{minipage}{0.475\hsize}
\begin{center}
\includegraphics[width=1\hsize,clip]{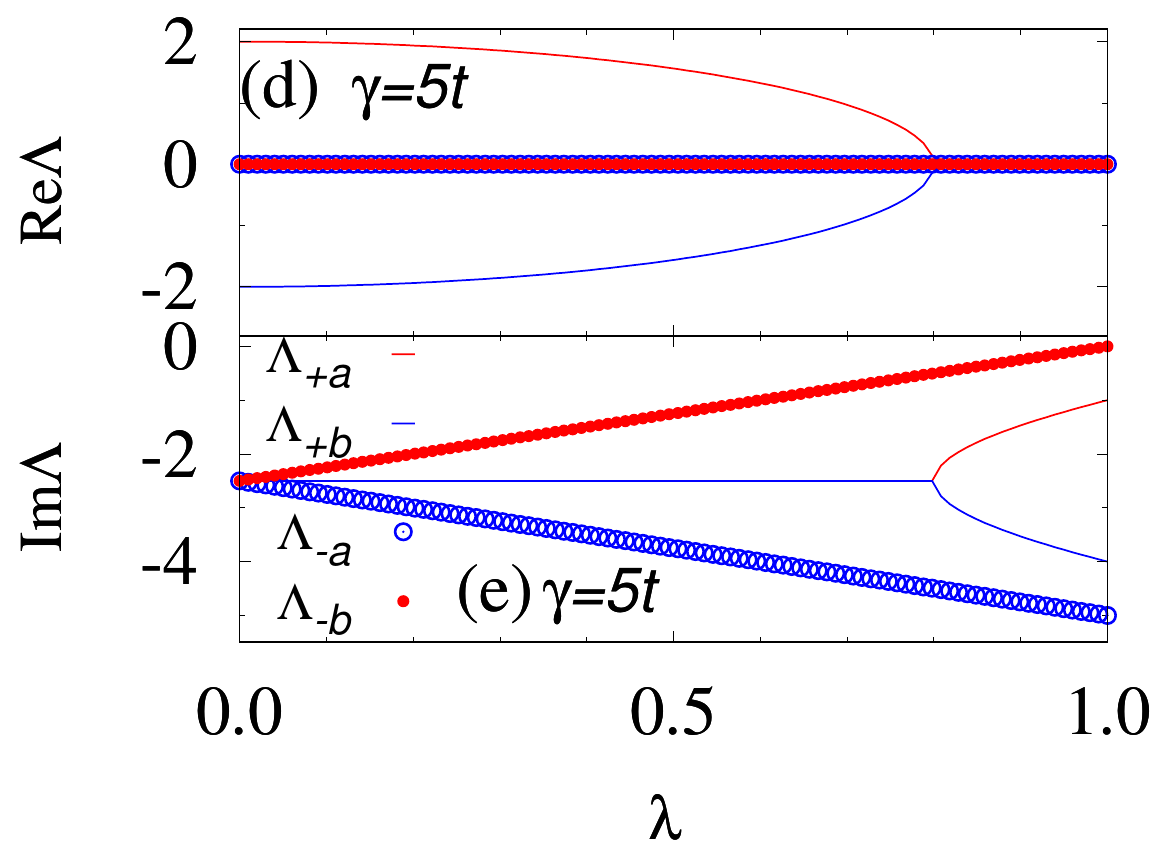}
\end{center}
\end{minipage}
\begin{minipage}{0.475\hsize}
\begin{center}
\includegraphics[width=1\hsize,clip]{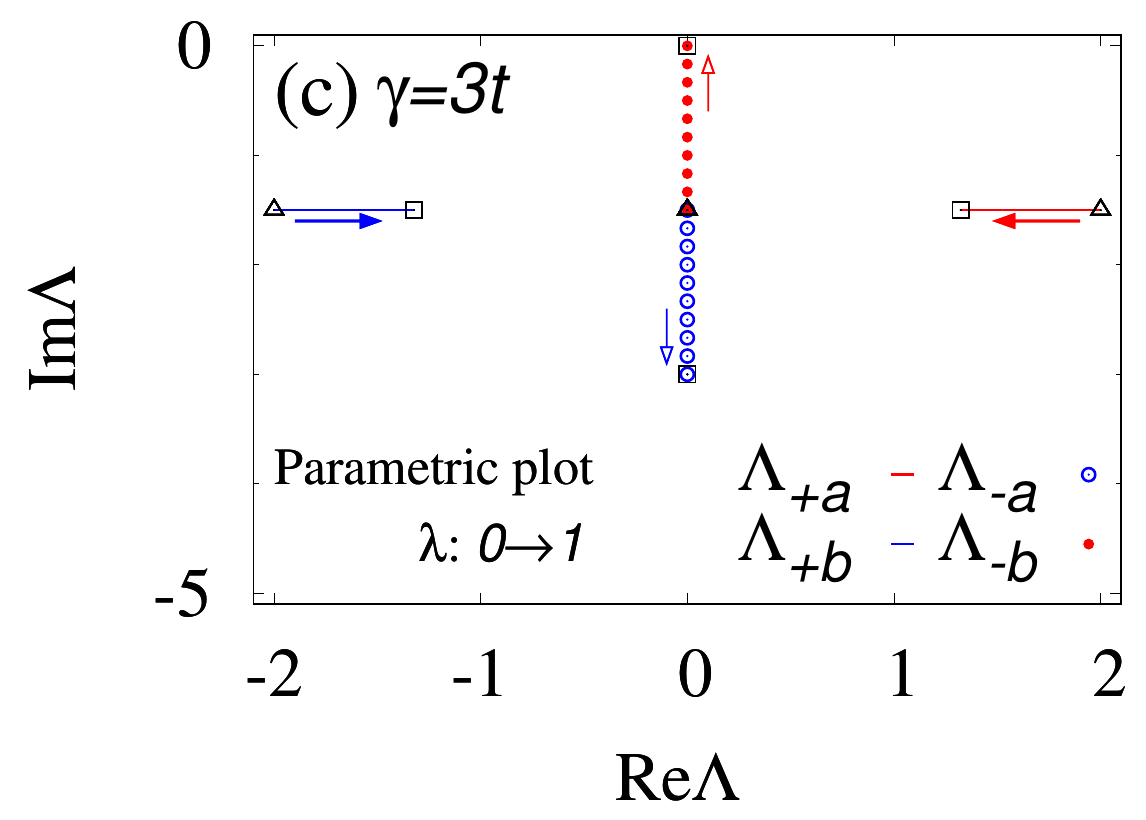}
\end{center}
\end{minipage}
\begin{minipage}{0.475\hsize}
\begin{center}
\includegraphics[width=1\hsize,clip]{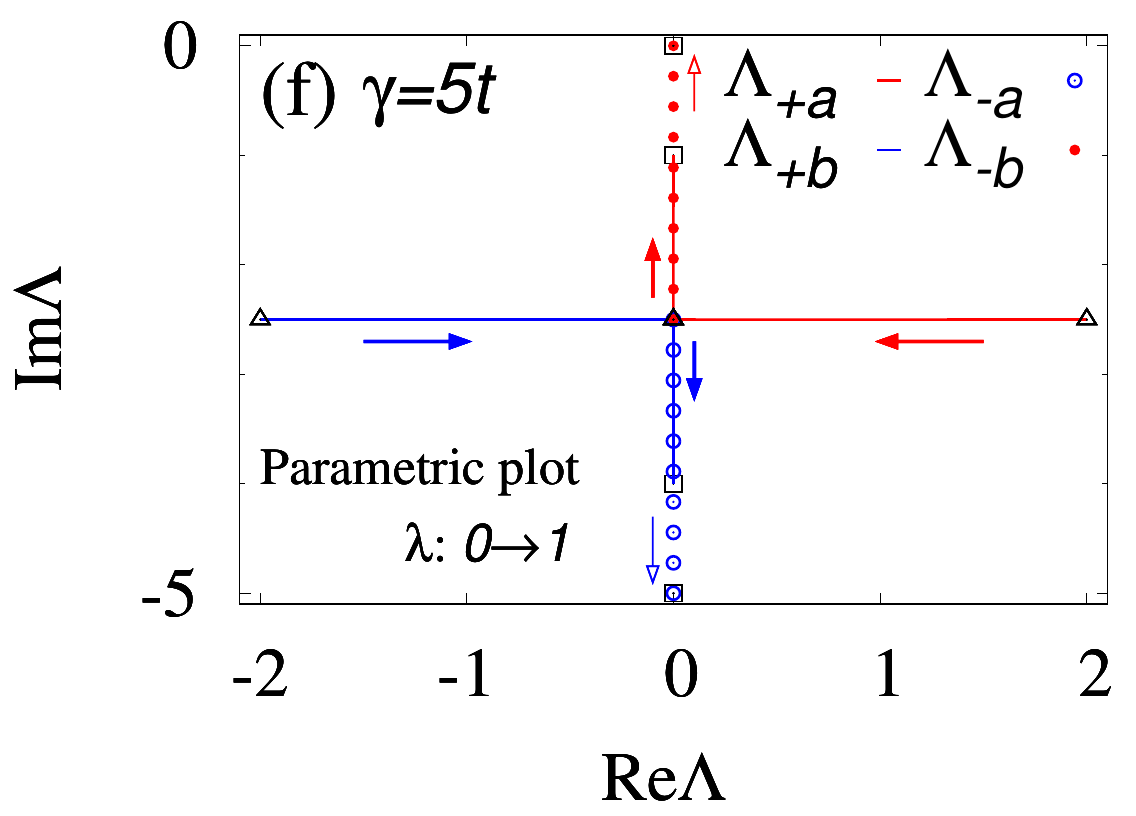}
\end{center}
\end{minipage}
\caption{(Color Online).
(a)-(c) [(d)-(f)]: The spectrum of the ``Liouvillian" $\matf{L}(\lambda)$ for $\gamma=3t$ and $(\gamma=5t)$. These data are obtained for $t=1$.
Panels (a) and (d) [(b) and (e)] show the real- (imaginary-) part of the eigenvalues as functions of $\lambda$. 
Panels (c) and (f) show the parametric plot of the spectrum for $0 \leq \lambda \leq 1$.
As $\lambda$ increases from $\lambda=0$ to $1$, the eigenvalues flow along the allows.
The data for $\lambda=0$ ($\lambda=1$) are plotted as triangles (squares).
We note that at $\lambda=1$, the exceptional point can be observed with increasing $\gamma$.
}
\label{fig: SSH Spec}
\end{figure}

In Fig.~\ref{fig: SSH Spec}, the spectrum of ``Liouvillian" $\matf{L}_{\mathrm{2site}}(\lambda)$ is plotted for $0\leq \lambda \leq 1$. For $4t\leq \gamma$, an exceptional point appears with increasing $\lambda$.
However, regardless of the value of $\gamma$, the eigenstate with eigenvalue $\Lambda_{-b}$ is the longest lifetime.
In particular, for $\lambda=1$, it is a non-equilibrium steady state, i.e., the lifetime become infinite.
From Eq.~(\ref{eq: 2site L matrix app}), we can see that the corresponding left and right eigenstates are ${}_L\llangle \rho_{2\mathrm{site},g} | = \llangle-2|$ and $| \rho_{2\mathrm{site},g} \rrangle_{R} = |-2\rrangle $.

For the state $|\rho_{2\mathrm{site},g}\rrangle_R$, the Berry phase takes $\pi$.
To see this, firstly, we note that twisting the hopping $t$ only for the subsector with $\sigma=\uparrow$ [see Eq.~(\ref{eq: L theta SSH gen app})] can be accomplished by applying the operator $e^{i\theta n_{1A\uparrow}}$~\cite{Hirano_gauge_PRB08};
\begin{eqnarray}
\label{eq: U L2site U app}
\matf{L}_{2\mathrm{site}}(\theta,\lambda) &=& e^{i\theta n_{1A\uparrow}} \matf{L}_{2\mathrm{site}}(\lambda) e^{-i\theta n_{1A\uparrow}}.
\end{eqnarray}
with $-\pi \leq \theta < \pi$.
Here, we note that Eq.~(\ref{eq: U L2site U app}) holds only for $t'=0$.
Equation~(\ref{eq: U L2site U app}) indicates that the eigenstates of $\matf{L}_{2\mathrm{site}}(\theta,\lambda)$ can be obtained from those of $\matf{L}_{2\mathrm{site}}(\lambda)$;
for instance, the eigenstate with the longest lifetime for $\matf{L}_{2\mathrm{site}}(\theta,\lambda)$ is given by
\begin{subequations}
\label{eq: SSH rhog_theta PBC}
\begin{eqnarray}
|\rho_{2\mathrm{site},g}(\theta)\rrangle_{R} &=& e^{i\theta n_{1A\uparrow}} |-2\rrangle, \\
{}_L\llangle \rho_{2\mathrm{site},g}(\theta)| &=&  \llangle -2 | e^{-i\theta n_{1A\uparrow}}.
\end{eqnarray}
\end{subequations}
Therefore, computing the eigenvalue of $I$, 
\begin{subequations}
\begin{eqnarray}
I| \rho_{2\mathrm{site},g}(0)\rrangle_{R}   &=& I \frac{1}{\sqrt{2}} \left(d^\dagger_{1A\uparrow}d^\dagger_{0B\downarrow} - d^\dagger_{0B\uparrow}d^\dagger_{1A\downarrow} \right) | 0 \rrangle \nonumber \\
                          &=&   \frac{1}{\sqrt{2}} \left(d^\dagger_{0B\uparrow}d^\dagger_{1A\downarrow} - d^\dagger_{1A\uparrow}d^\dagger_{0B\downarrow} \right) | 0 \rrangle \nonumber \\
                          &=&  -|\rho_{2\mathrm{site},g}(0)\rrangle_{R}, \\
I|\rho_{2\mathrm{site},g}(\pi)\rrangle_{R} &=& I \frac{1}{\sqrt{2}} \left(-d^\dagger_{1A\uparrow}d^\dagger_{0B\downarrow} - d^\dagger_{0B\uparrow}d^\dagger_{1A\downarrow} \right) | 0 \rrangle \nonumber \\
                          &=&   \frac{1}{\sqrt{2}} \left(-d^\dagger_{0B\uparrow}d^\dagger_{1A\downarrow} - d^\dagger_{1A\uparrow}d^\dagger_{0B\downarrow} \right) | 0 \rrangle \nonumber \\
                          &=& |\rho_{2\mathrm{site},g}(\pi)\rrangle_{R},
\end{eqnarray}
\end{subequations}
yields the Berry phase $\chi_{\uparrow g}=\pi$. Here, we have used Eq.~(\ref{eq: chi Z2 app}).
We note that the same result can be obtained by direct evaluation of the integral in Eq.~(\ref{eq: def of chi_k app})~\cite{Berry_SSH_dephasing_ftnt}.

Corresponding to the Berry phase taking $\pi$, one may expect the emergence of edge states~\cite{Ryu_ZeroGra_PRL02,Hatsugai_BulkEdge_09} which is discussed at the end of this section. 
Here, for comparison, we discuss expectation values under the periodic boundary condition.
Firstly, we note that the state is written as
\begin{eqnarray}
\label{eq: |-2> to rho_g app}
|\rho_{2\mathrm{site},g} \rrangle &\to&  \rho_{2\mathrm{site},g} =\frac{1}{2} \left( c^\dagger_{1A} |0\rangle \langle 0| c_{1A} + c^\dagger_{0B} |0\rangle \langle 0| c_{0B} \right), \nonumber \\
\end{eqnarray}
which we see below.
Here, we have normalized the density matrix so that $\mathrm{tr}\rho_{2\mathrm{site},g}=1$ holds.
Thus, we obtain
\begin{eqnarray}
\label{eq: SSH <n> PBC app}
\mathrm{tr} (n_{1A} \rho_{2\mathrm{site},g}) =\frac{1}{2},  &&\quad  \mathrm{tr} (n_{0B} \rho_{2\mathrm{site},g}) =\frac{1}{2}. 
\end{eqnarray}
Equation~(\ref{eq: |-2> to rho_g app}) can be seen by a straightforward calculation.
As we have applied the particle-hole transformation [see Eq.~(\ref{eq: particle-hole SSH app})], $|\rho_{2\mathrm{site},g}\rrangle =|-2\rrangle$ is mapped as
\begin{eqnarray}
|-2\rrangle &\to& \frac{1}{ \sqrt{2} } \left(d^\dagger_{1A\uparrow} d_{0B\downarrow} -d^\dagger_{0B\uparrow} d_{1A\downarrow} \right) d^\dagger_{1A\downarrow}d^\dagger_{0B\downarrow}|0 \rrangle \nonumber  \\
           &=& -\frac{1}{ \sqrt{2}  } \left(d^\dagger_{1A\uparrow} d^\dagger_{1A\downarrow} +d^\dagger_{0B\uparrow} d^\dagger_{0B\downarrow} \right) |0\rrangle, 
\end{eqnarray}
which can be rewritten in terms of $c_{i\alpha}$ and $\bar{c}_{i\alpha}$ as follows:
\begin{eqnarray}
|-2 \rrangle  &\to& \frac{1}{ \sqrt{2} } P_{fc} \left(c^\dagger_{1A} \bar{c}^\dagger_{1A} +c^\dagger_{0B} \bar{c}^\dagger_{0B} \right) |0 \rrangle,
\end{eqnarray}
where $P_{fc}:=(-1)^{c^\dagger_{1A} c_{1A} +c^\dagger_{0B} c_{0B} }$.
By normalizing the density matrix so that $\mathrm{tr}(\rho_{2\mathrm{site},g})=1$ holds, we obtain Eq.~(\ref{eq: |-2> to rho_g app}).
In the above, we have seen that Eq.~(\ref{eq: SSH <n> PBC app}) holds for the periodic boundary condition.

\textit{(ii) Edge properties--.}
Now, let us analyze the system with edges. We impose the open boundary condition; sites $i=0$ and $i=L-1$ are decoupled.
We again restrict ourselves to the half-filled case.
For $t'=0$, each boundary site is isolated from the bulk.
The ``Liouvillian" at the edge $j=0$ is written as $\matf{L}_{\mathrm{edge}}(\lambda)=-i\lambda \gamma(n_{0A\uparrow}-\frac{1}{2})(n_{0A\downarrow}-\frac{1}{2})-\frac{i\gamma}{4}$.
The right eigenvectors and corresponding eigenvalues are easily obtained and written as
\begin{subequations}
\begin{eqnarray}
               |0\rrangle, & & \quad \Lambda_{0} = -\frac{i\gamma}{4}(1+\lambda), \\
d^\dagger_{0A\uparrow} |0\rrangle, & & \quad \Lambda_{\uparrow} = -\frac{i\gamma}{4}(1-\lambda), \\
d^\dagger_{0A\downarrow} |0\rrangle, & & \quad \Lambda_{\downarrow} = -\frac{i\gamma}{4}(1-\lambda), \\
d^\dagger_{0A\uparrow}d^\dagger_{0A\downarrow} |0\rrangle, & & \quad \Lambda_{\uparrow\downarrow} = -\frac{i\gamma}{4}(1+\lambda).
\end{eqnarray}
\end{subequations}
%
%
Here, we note that the states with the longest lifetime are doubly degenerate.
Taking into account two edges, we obtain the edge state with an infinite lifetime, 
\begin{eqnarray}
|\rho_{\mathrm{edge},g} \rrangle &=& \left(a d^\dagger_{0A\uparrow}d^\dagger_{L-1B\downarrow} +b d^\dagger_{L-1B\uparrow}d^\dagger_{0A\downarrow} \right) |0\rrangle, \nonumber \\
\end{eqnarray}
with real numbers $a$ and $b$ satisfying $a^2+b^2=1$.
We note that $d^\dagger_{0A\uparrow}d^\dagger_{L-1B\uparrow} |0\rrangle$ is also an eigenstate with the zero eigenvalue. 
However, we discard this states because we restrict ourselves to the half-filled case, $(N_\uparrow,N_\downarrow)=(1,1)$ with $N_{\sigma}=n_{0A\sigma}+n_{L-1B\sigma}$.

As shown below, $|\rho_{\mathrm{edge},g}\rrangle$ can be rewritten as
\begin{eqnarray}
\label{eq: |rho_g> to rho_g OBC app}
\!\!\!\!\!\!\!\!\!\!\!\! | \rho_{\mathrm{edge},g} \rrangle &\to& \rho_{\mathrm{edge},g} = \left( a' c^\dagger_{0A} |0\rangle\langle0|  c_{0A}  \right. \nonumber \\
 && \qqqqquad \left. - b' c^\dagger_{L-1B} |0\rangle\langle0| c_{L-1B} \right), 
\end{eqnarray}
with $a'$ and $b'$  are real numbers satisfying $a'-b'=1$.
Here, we have renormalized the states so that $\mathrm{tr}\left(\rho_{\mathrm{edge},g}\right)=1$ holds.
Therefore, we obtain
\begin{eqnarray}
 \mathrm{tr}(n_{0A} \rho_{\mathrm{edge},g} ) = a',  &&\quad \mathrm{tr}(n_{L-1B} \rho_{\mathrm{edge},g} ) = -b'.
\end{eqnarray}

This result means that the polarization is observed only at each edge. Namely, we have
\begin{subequations}
\begin{eqnarray}
\mathrm{tr}\left[ (n_{0A}-n_{0B} ) \rho_{\mathrm{edge},g} \right] &=& a'-\frac{1}{2},
\end{eqnarray}
at $j=0$, while we have
\begin{eqnarray}
\mathrm{tr}\left[ (n_{jA}-n_{jB})\rho_{\mathrm{edge},g} \right]&=&0,
\end{eqnarray}
\end{subequations}
for the bulk ($j=1,\cdots,L-2$) [see Eq.~(\ref{eq: SSH <n> PBC app})].

Equation~(\ref{eq: |rho_g> to rho_g OBC app}) can be obtained in a similar way to the analysis of the bulk [see Eq.~(\ref{eq: |-2> to rho_g app})].
As we have applied the particle-hole transformation [see Eq.~(\ref{eq: particle-hole SSH app})] the state $| \rho_{\mathrm{edge},g} \rrangle$ is mapped as
\begin{eqnarray}
| \rho_{\mathrm{edge},g} \rrangle & \to& \left(a d^\dagger_{0A\uparrow}d_{L-1B\downarrow} +b d^\dagger_{L-1B\uparrow}d_{0A\downarrow} \right) d^\dagger_{0A\downarrow}d^\dagger_{L-1B\downarrow} |0\rrangle \nonumber \\
                                            &=& \left(-a d^\dagger_{0A\uparrow}d^\dagger_{0A\downarrow} +b d^\dagger_{L-1B\uparrow}d^\dagger_{L-1B\downarrow} \right)  |0\rrangle, \nonumber \\
\end{eqnarray}
which can be rewritten in terms of $c_{i\alpha}$ and $\bar{c}_{i\alpha}$ as follows:
\begin{eqnarray}
| \rho_{\mathrm{edge},g} \rrangle  &\to& P_{fc} ( a c^\dagger_{0A}\bar{c}^\dagger_{0A} - b c^\dagger_{L-1B} \bar{c}^\dagger_{L-1B} ) |0\rrangle,
\end{eqnarray}
where $P_{fc}:=(-1)^{ c^\dagger_{0A}c_{0A}+c^\dagger_{L-1B}c_{L-1B} }$.
By normalizing the density matrix so that $\mathrm{tr}(\rho_{\mathrm{edge},g})=1$ holds, we obtain Eq.~(\ref{eq: |rho_g> to rho_g OBC app}).

In the above, for $t'=0$, the Berry phase $\chi_{\uparrow g}$ of the non-equilibrium steady states takes $\pi$. 
Correspondingly, while the charge distribution of the bulk is uniform, each edge shows the charge polarization.

We recall that the topological properties remain unchanged as long as the gap does not close. This fact means that for small but finite $t'$, the Berry phase should take $\pi$ inducing the edge polarization.

\section{
Spectrum of a block-upper-triangular matrix
}
\label{sec: triangular app}

The spectrum of the ``Liouvillian" $\matf{L}(\lambda)=\matf{L}_0+\lambda \matf{L}_{\mathrm{J}}$ is independent of $\lambda$ ($0 \leq  \lambda  \leq 1$) when $\matf{L}_{\mathrm{J}}$ ($\matf{L}_0$) is a block-upper-triangular (block-diagonal) matrix~\cite{Torres_Utrian_PRA14,Nakagawa_exactHubb_arXiv20}.

In order to see this, let us consider the following square matrix of a block-upper-triangular form, 
\begin{eqnarray}
\label{eq: L0to4 upper eg app}
\matf{L}(\lambda)
&=&
\left(
\begin{array}{ccc}
\mathcal{L}_{(0,0)} & \lambda \mathcal{L}_{\mathrm{J}(0,2)}  & 0 \\
0 & \mathcal{L}_{(2,2)} & \lambda \mathcal{L}_{\mathrm{J}(2,4)} \\
0  & 0 & \mathcal{L}_{(4,4)} 
\end{array}
\right),
\end{eqnarray}
where
$\mathcal{L}_{(0,0)}$, $\mathcal{L}_{(2,2)}$, and $\mathcal{L}_{(4,4)}$ are non-Hermitian square matrices.
Matrices $\mathcal{L}_{\mathrm{J}(0,2)}$ and $\mathcal{L}_{\mathrm{J}(2,4)}$ are non-Hermitian and not necessarily square matrices.
The spectrum of $\matf{L}(\lambda)$ is independent of $\lambda$, which can be seen as follows.

Firstly, we note that an arbitrary eigenvalue $\Lambda$ of $\matf{L}(\lambda)$ in Eq.~(\ref{eq: L0to4 upper eg app}) is determined by the characteristic equation,
\begin{eqnarray}
\mathrm{det}\left[
\left(
\begin{array}{ccc}
\mathcal{L}_{(0,0)} & \lambda \mathcal{L}_{\mathrm{J}(0,2)}  & 0 \\
0 & \mathcal{L}_{(2,2)} & \lambda \mathcal{L}_{\mathrm{J}(2,4)} \\
0  & 0 & \mathcal{L}_{(4,4)} 
\end{array}
\right)
-\Lambda \1
\right]
&=& 0.
\end{eqnarray}
%
Regardless of the value of $\lambda$, the above equation is rewritten as~\cite{uptri_ftnt} $\mathrm{det}(\mathcal{L}_{(0,0)}-\Lambda\1)\mathrm{det}(\mathcal{L}_{(2,2)}-\Lambda\1)\mathrm{det}(\mathcal{L}_{(4,4)}-\Lambda\1)=0$, which indicates that the spectrum of the matrix $\matf{L}(\lambda)$ is independent of $\lambda$.

The above argument can be straightforwardly extended to a generic case.
Thus, we can conclude that  the spectrum of the ``Liouvillian" $\matf{L}(\lambda)= \matf{L}_0+\lambda \matf{L}_{\mathrm{J}}$ is independent of $\lambda$ when $\matf{L}_{\mathrm{J}}$ ($\matf{L}_0$) is a block-upper-triangular (block-diagonal) matrix.

\section{
Quantization of the pseudo-spin Chern number
}
\label{sec: quantize ps Ch app}
The pseudo-spin Chern number is quantized even in the presence of the jump term. 
To see this, we show that $C_{\sigma\sigma}$ ($\sigma=K,B$) defined in Eq.~(\ref{eq: Ckk w jump}) is quantized.
\blue{
We note that the quantization of a many-body Chern number for non-Hermitian systems is proven~\cite{Yoshida_nHFQH19} by extending the proof in the Hermitian case~\cite{Niu_HallCond_PRB85,Kohmoto_AnnPhys1985}.
We note, however, that, quantization of the non-Hermitian many-body Chern numbers ($C_{KK}$ and $C_{BB}$), which are computed by twisting the boundary condition only for a subsector of the Hilbert space, has not been proven yet. Thus, this section is devoted to its proof.
}

Consider ``Liouvillian" $\matf{L}(\theta_x,\theta_y,\lambda)$ with $0\leq \theta_{x(y)}<2\pi$ and $ 0 \leq \lambda \leq 1$ which is obtained by twisting the boundary condition only for the subspace specified by $\sigma$.
Because taking the unique gauge may not be allowed, we divide the two-dimensional space $(\theta_x,\theta_y)$ into two regions, I and I\!I so that the eigenstates are single-valued and are smoothly defined in each region. We note in passing that one can treat the case, where the space $(\theta_x, \theta_y)$ needs to be divided into more than three regions, on an equal footing.

In each region, the Berry curvature is rewritten as
\begin{subequations}
\begin{eqnarray}
F_{\sigma\sigma}&=&\partial^\sigma_x A^s_{\sigma y}-\partial^\sigma_y A^s_{\sigma x}, \\
A^s_{\sigma\mu}&:=&\sum_n {}_L \llangle  \rho^s_n | \partial^\sigma_\mu | \rho^s_n \rrangle_R,
\end{eqnarray}
\end{subequations}
with $\mu=x,y$.
Here, $| \rho^s_n \rrangle_R$ and ${}_L \llangle  \rho^s_n |$ are right and left eigenstates of $\matf{L}(\theta_x,\theta_y,\lambda)$ for region $s=\I,\II$.
The summation $\displaystyle{\sum_n}$ is taken over degenerate states.
By making use of Stokes' theorem, $C_{\sigma\sigma}$ defined in Eq.~(\ref{eq: Ckk w jump}a) can be written as
\begin{eqnarray}
\label{eq: Csigsig in Z app}
C_{\sigma\sigma} &=& \frac{1}{2\pi }\oint \!\! d\theta_\mu \, \mathrm{Im} (A^{\I}_{\sigma\mu}-A^{\II}_{\sigma\mu}) \nonumber \\
                 &=&\frac{1}{2\pi}\oint \!\! d\theta_\mu \, \mathrm{Im} \partial_\mu \log(\mathrm{det}M) \in \mathbb{Z}.
\end{eqnarray}
Here, the integral is taken over the boundary of two regions, $\I$ and $\II$, and $M$ is an invertible matrix.
From the first to the second line, we have used the following relation
\begin{eqnarray}
A^{\I}_{\sigma \mu} &=& A^{\II}_{\sigma \mu}+\sum_{nm} M^{-1}_{nm} \partial^{\sigma}_\mu M_{mn}.
\end{eqnarray}
This relation is obtained by noting that relations $|\rho^{\I}_n \rrangle_R =\sum_m |\rho^{\II}_m \rrangle_R M_{mn}$ and  $ {}_L \llangle \rho^{\I}_n | =\sum_m  M^{-1}_{nm} {}_L \llangle \rho^{\II}_m |$ hold because both of the gauges are available on the boundary of two regions $\I$ and $\II$.
We recall that the biorthogonal normalization condition is imposed on the right and left eigenvectors.

Equation~(\ref{eq: Csigsig in Z app}) indicates the quantization of $C_{\sigma\sigma}$.
We note that Eq.~(\ref{eq: Csigsig in Z app}) holds as long as the gap-closing does not occur in the parameter space $(\theta_x,\theta_y)$.

%
\blue{
We note that introducing a perturbation does not change $C_{\sigma\sigma}$ as long as the gap is open~\cite{Katsura_disorder_Ch_JMP16}. 
This is because $C_{\sigma\sigma}$ is continuous as a function of the strength of the perturbation maintaining the gap, while $C_{\sigma\sigma}$ is quantized [see Eq.~(\ref{eq: Csigsig in Z app})].
}

\blue{
We close this section by noting 
}
that Eq.~(\ref{eq: Ckk w jump}a) is written as
$
C_{\sigma\sigma} = \frac{1}{ 2\pi i }\int \! d\theta_xd\theta_y \, F_{\sigma\sigma}.
$
This is because the integral of the real-part of the Berry curvature vanishes; the real-part of a complex function $\log z$ with $z \in \mathbb{C}$ is single-valued.

\section{
Liouvillian with the pseudo-potential approximation
}
\label{sec: pp Liou app}
Here, with the pseudo-potential approximation, we see that the Liouvillian~(\ref{eq: C-iso L nHFQH}) can be written as
\begin{subequations}
\begin{eqnarray}
\label{eq: pp Liou app}
\matf{L} &\simeq& \sum_{ij\sigma} h_{ij\sigma} f^\dagger_{i\sigma}f_{j\sigma} +\sum_{\langle ij\rangle\sigma} V_{\sigma} f^\dagger_{i\sigma} f^\dagger_{j\sigma} f_{j\sigma} f_{i\sigma} \nonumber \\
&&-i\gamma\sum_{\langle ij \rangle} f_{ia}f_{ja} f_{jb} f_{ib},
\end{eqnarray}
where
\begin{eqnarray}
f^\dagger_{ia}:=\displaystyle{\sum}'_{n_1} \varphi^*_{in_1} a^\dagger_{n_1a}, &\quad& f^\dagger_{ib}:=\displaystyle{\sum}'_{n_1} \varphi_{in_1} a^\dagger_{n_1b}.
\end{eqnarray}
\end{subequations}
Here, $h_{ij\sigma}$ and $V_\sigma$ are defined just below Eq.~(\ref{eq: C-iso L nHFQH}).
The operator $a^\dagger_{n_1\sigma}$ creates the fermion in state $\varphi_{in_1}$ of the lowest Landau level for layer $\sigma$ ($\sigma=a,b$). 
The creation and the annihilation operators satisfy $\{ a _{n_1\sigma}, a^\dagger_{n_2\sigma'} \} =\delta_{n_1n_2}\delta_{\sigma\sigma'}$ and $\{ a_{n_1\sigma}, a_{n_2\sigma'} \}=0$.
The summation $\sum'_{n_1}$ is taken over the states in the lowest Landau level.

In the following, we derive Eq.~(\ref{eq: pp Liou app}).
Firstly, we note that the anti-commutation relation between $a^\dagger_{n_1a}$ and $a^\dagger_{n_1b}$ is due to the introduction of the operator for the fermion number parity. 
Namely, we can see that $\rho a^\dagger_{n_1}$ is mapped as  $\bar{a}_{n_1} | \rho \rrangle$. 
The annihilation operator $\bar{a}_{n_1}$ acts on a state in subspace $\mathrm{Bra}$. 
The operators $\bar{a}$'s commute with the operators $a$'s and $a^\dagger$'s, $[\bar{a}_{n_1}, a^\dagger_{n_2}]=[\bar{a}_{n_1}, a _{n_2}]=0$.
Thus, introducing operators $a^\dagger_{n_1a}:= a^\dagger_{n_1}$ and $a^\dagger_{n_1b}:= \bar{a}^\dagger_{n_1}P_{fa}$ ($P_{fa}=(-1)^{\sum'_{n_1} a^\dagger_{n_1a} a _{n_1a} }$), we have the anti-commutation relation between $a^\dagger_{n_1a}$ and  $a^\dagger_{n_2b}$.

Secondly, we note that with the pseudo-potential approximation, the operators can be written as follows:
\begin{subequations}
\begin{eqnarray}
H_{0}&\simeq& \sum_{ij} h_{ij} f^\dagger_i f_j +V_R\sum_{\langle ij\rangle} f^\dagger_i f^\dagger_j f_j f_i, \nonumber \\
\sum_{\alpha} L^{\dagger}_\alpha L_\alpha &=& \gamma \sum_{\langle ij\rangle} c^\dagger_jc^\dagger_ic_ic_j \simeq  \gamma \sum_{\langle ij\rangle} f^\dagger_if^\dagger_jf_jf_i, \nonumber \\
\sum_{\alpha} L_\alpha \rho L^\dagger_\alpha  &=& \gamma \sum_{\langle ij \rangle} c_i c_j \rho c^\dagger_jc^\dagger_i \simeq \gamma \sum_{\langle ij \rangle} f_i f_j \rho f^\dagger_jf^\dagger_i. \nonumber
\end{eqnarray}
\end{subequations}

With the isomorphism [see Eq.~(\ref{eq: choi iso rho})], these terms can be identified as follows:
\begin{subequations}
\begin{eqnarray}
 \rho h_{ij}(f^\dagger_i f_j) &\leftrightarrow & h_{ij} f^\dagger_{jb} f^\dagger_{ib} |\rho \rrangle = (h^*_{ji}) f^\dagger_{jb} f^\dagger_{ib} |\rho \rrangle, \nonumber \\
 \rho f^\dagger_if^\dagger_jf_jf_i  & \leftrightarrow & f^\dagger_{ib} f^\dagger_{jb} f_{jb} f_{ib} |\rho \rrangle, \nonumber \\
f_i f_j \rho f^\dagger_jf^\dagger_i & \leftrightarrow &   f_{ia} f_{ja}  f_{ib} f_{jb} |\rho\rrangle = -f_{ia} f_{ja} f_{jb}f_{ib} |\rho\rrangle. \nonumber 
\end{eqnarray}
\end{subequations}
%
Here, we have assumed $i\neq j$.
By taking into account the above relations, we get Eq.~(\ref{eq: pp Liou app}).

\section{
Topological degeneracy for another type of dissipation
}
\label{sec: topo deg jump app}

By a topological argument, we show that the system with the filling factor $\nu$ ($\nu^{-1}=1,3,5,\cdots$) shows at least $\nu^{-1}$-fold topological degeneracy in the spectrum of the Liouvillian when the Lindblad operators preserve charge $\mathrm{U}(1)$ symmetry.
We consider fermions in the square lattice (see Fig.~\ref{fig: model}) with $L_x=L_y=L$ and $\phi=1/L$. In this case, the number of states in the lowest Landau level is $N_{\phi}=L=\phi^{-1}$ (i.e., the filling factor is $\nu:=N_a/N_\phi= \phi N_a$).

Firstly, let us consider the eigenvectors of the kinetic terms under the Landau gauge:
\begin{eqnarray}
\sum_{j} h_{ij} \varphi_{jn_1}(k_y) &=& \varphi_{in_1}(k_y)\epsilon_{n_1},
\end{eqnarray}
with $n_1=1,\cdots, \mathrm{dim}\, h$.
Here, we note that the Hamiltonian $h_{ij}$ is invariant under the translation along the $y$-direction, meaning that the Landau state $\varphi_{jn_1}$ can also be labeled by momentum along the $y$-direction $k_y$:
\begin{eqnarray}
\label{eq: def of T_y |k> app}
T_y |\varphi_{n_1}(k_y)\rangle &=& e^{-ik_y}|\varphi_{n_1}(k_y)\rangle,
\end{eqnarray}
with $T_y$ being the translation operator along the $y$-direction.
In addition, for $L_x=L_y=L$ and $\phi=1/L$, the following relation holds~\cite{3fold_Liou_ftnt}:
\begin{subequations}
\label{eq: U |k> = |k+dK> app}
\begin{eqnarray}
U |\varphi_{n_1}(k_y)\rangle  = |\varphi_{n_1}(k_y-2\pi\phi)\rangle,
\end{eqnarray}
with 
\begin{eqnarray}
U c^\dagger_{j_xj_y} U^\dagger &=& e^{-2\pi i \phi j_y} c^\dagger_{j_xj_y}.
\end{eqnarray}
\end{subequations}
%
With the isomorphism [see Eq.~(\ref{eq: choi iso rho})], we obtain the following relations corresponding to Eqs.~(\ref{eq: def of T_y |k> app})~and~(\ref{eq: U |k> = |k+dK> app}): 
\begin{eqnarray}
 T_{y\sigma} | \varphi_{n_1} (k_y) \rrangle_\sigma &=& e^{-i\sgn(\sigma)k_y } | \varphi_{n_1} (k_y) \rrangle_\sigma, 
\end{eqnarray}
and 
\begin{subequations}
\begin{eqnarray}
\label{eq: U |k> = |k+dK> iso app}
 U_\sigma |\varphi_{n_1}(k_y)\rrangle_{\sigma}  &=& |\varphi_{n_1}(k_y-2\pi\phi)\rrangle_\sigma, \\
 U_\sigma d^\dagger_{j_xj_y\sigma} U^\dagger_\sigma  &=&  e^{- 2\pi i \phi \sgn(\sigma) j_y} d^\dagger_{j_xj_y\sigma}.
\end{eqnarray}
\end{subequations}
Here, $\sigma=a$ ($\sigma=b$) specifies the subspace $\mathrm{Ket}$ ($\mathrm{Bra}$).
The operator $d^\dagger_{j_xj_y\sigma}$ is the creation operator defined in Eq.~(\ref{eq: def of dia and dib}) where the set of the subscripts $j_x$ and $j_y$ is denoted by $j$.
Here, $T_{y\sigma}$ and $U_\sigma$ are defined as
\begin{eqnarray}
T_{ya}=T_y \otimes \1, &\quad\quad& T_{yb}=\1 \otimes T^*_y, \\
U_{a} =U \otimes \1,  &\quad\quad& U_{a}=\1 \otimes U^*.
\end{eqnarray}

With the above relation, we can see that the system shows robust topological degeneracy when the following conditions are satisfied:
\begin{eqnarray}
\label{eq: U1 cons Lmu app}
U_{\sigma} L_{\alpha \sigma} U^\dagger_{\sigma} &=& L_{\alpha \sigma}, \\
\label{eq: trans Liou app}
T_{ya}T_{yb}\matf{L}(T_{ya}T_{yb})^\dagger &=& \matf{L}.
\end{eqnarray}
with $L_{\alpha a}=L_\alpha\otimes \1$ and $L_{\alpha b}=\1 \otimes L^T_\alpha$.

To see the robust topological degeneracy, firstly, we note that the Liouvillian can be block-diagonalized into sectors each of which is labeled by the momentum $K_y$ and the number of fermions.
By making use of Eq.~(\ref{eq: U |k> = |k+dK> iso app}), we can see the relation between the matrices for each sector
\begin{eqnarray}
\label{eq: L mat ele trans app}
&& \llangle \Phi_{\{n\}}(K_{y}) | \matf{L} |\Phi_{\{n'\}}(K'_{y}) \rrangle \nonumber \\
&&= \llangle \Phi_{\{n\}}(K_{y}) | U^\dagger_a \matf{L}  U_a |\Phi_{\{n'\}}(K'_{y}) \rrangle \nonumber \\
&&= \llangle \Phi_{\{n\}}(K_{y}+\Delta K) | \matf{L}  |\Phi_{\{n'\}}(K'_{y}+\Delta K) \rrangle. \nonumber \\
\end{eqnarray}
Here, $| \Psi_{\{n_\sigma\}} (K_{y_\sigma}) \rrangle_\sigma$ is defined as
\begin{eqnarray}
|\Phi_{\{n\}}(K_{y}) \rrangle 
&=&
| \Phi_{\{n_a\}} (K_{ya}) \rrangle_a \otimes | \Phi_{\{n_b\}} (K_{yb}) \rrangle_b, \nonumber \\
\end{eqnarray}
\begin{eqnarray}
| \Phi_{\{n_\sigma\}} (K_{y\sigma}) \rrangle_\sigma 
 &=& 
 a^\dagger_{n_1k_1\sigma}a^\dagger_{n_2k_2\sigma}\cdots a^\dagger_{n_N k_N \sigma} |0\rrangle_\sigma,\nonumber \\
\end{eqnarray}
with $K_y=K_{ya}-K_{yb}$ and $K_{y\sigma}=\sum_{l=1,\cdots,N_\sigma} k_{yl}$.
The operator $a^\dagger_{n_lk_{l'}\sigma}$ with $l,l'=1,2,\cdots$ creates a fermion in state $|\varphi_{n_{l} }(k_{yl'})\rrangle_\sigma$ at layer $\sigma=a,b$ (see footnote~\onlinecite{defs_adag_a_ftnt}), and $|0\rrangle_\sigma$ is the vacuum being the state annihilated by all $a_{n_{l}\sigma}$.

By noting the relation $\Delta K =-2\pi \phi N_a =-2\pi \nu$ for $L_x=L_y=L$ and $\phi=1/L$, we see that: for $\nu^{-1}=1,3,5,\cdots$, the Liouvillian $\matf{L}$ can be block-diagonalized into $\nu^{-1}$ subsectors labeled by momentum $K_y$ [see Eq.~(\ref{eq: trans Liou app})]; these block-diagonalized matrices are identical to each other [see Eq.~(\ref{eq: L mat ele trans app})].

Therefore, we can conclude that regardless of details of the dissipation, the open quantum system shows at least $\nu^{-1}$-fold degeneracy as long as 
both $\mathrm{U}(1)$ symmetry [Eq.~(\ref{eq: U1 cons Lmu app})] and translational symmetry [Eq.~(\ref{eq: trans Liou app})] are preserved.
Namely, in the absence of accidental degeneracy, we have $\nu^{-1}$-fold degeneracy which is topologically protected.

\end{document}